\documentclass[apj]{emulateapj}
\usepackage{apjfonts}

\def\ltsim{\raise 2pt \hbox {$<$} \kern-0.85em \lower 3pt \hbox {$\sim$}}
\def\gtsim{\raise 2pt \hbox {$>$} \kern-0.85em \lower 3pt \hbox {$\sim$}}

\slugcomment{Received 2010 November 8; revised 2011 February 4; accepted 2011 March 5}

\journalinfo{The Astrophysical Journal Supplement Series, in press}

\shorttitle{Radio/X-ray study of galaxy groups}
\shortauthors{GIACINTUCCI ET AL.}
\begin{document}

\title{A combined low-radio frequency/X-ray study of galaxy groups\\  I. {\em Giant Metrewave Radio Telescope} observations 
at 235 MHz and 610 MHz}

\author{Simona Giacintucci\altaffilmark{1,2,3}, 
Ewan O'Sullivan\altaffilmark{2,4},
Jan Vrtilek\altaffilmark{2}, 
Laurence P. David\altaffilmark{2},
Somak Raychaudhury\altaffilmark{4},
Tiziana Venturi\altaffilmark{3}, 
Ramana M. Athreya\altaffilmark{5}, 
Tracy E. Clarke\altaffilmark{6},
Matteo Murgia\altaffilmark{7}, 
Pasquale Mazzotta\altaffilmark{2,8},
Myriam Gitti\altaffilmark{2,9},
Trevor Ponman\altaffilmark{4},
C.~H.~Ishwara-Chandra\altaffilmark{10},
Christine Jones\altaffilmark{2}, 
William R. Forman\altaffilmark{2}}

\altaffiltext{1}{Department of Astronomy, University of Maryland,
  College Park, MD 20742-2421; simona@astro.umd.edu}
\altaffiltext{2}{Harvard-Smithsonian Center for Astrophysics, 
                 60 Garden Street, Cambridge, MA 02138, USA}
\altaffiltext{3}{ INAF - Istituto di Radioastronomia, via Gobetti 101, 
                  I-40129, Bologna, Italy} 
\altaffiltext{4}{School of Physics and Astronomy, University of Birmingham,
Edgbaston, Birmingham B15 2TT - UK}
\altaffiltext{5}{IISER, Pune, Maharashtra, 411008, India}
\altaffiltext{6}{Naval Research Laboratory, 4555 Overlook Ave. SW,
  Code 7213, Washington, DC 20375, USA}
\altaffiltext{7}{INAF - Osservatorio Astronomico di Cagliari,
  Loc. Poggio dei Pini, Strada 54, I-09012 Capoterra (CA), Italy}
\altaffiltext{8}{Dipartimento di Fisica, Universit\'a di Roma Tor
  Vergata, via della Ricerca Scientifica 1, I-00133, Roma, Italy}
\altaffiltext{9}{Osservatorio Astronomico di Bologna - INAF,
via Ranzani 1, I-40127 Bologna  - Italy}
\altaffiltext{10}{National Centre for Radio Astrophysics, TIFR, Post
  Bag No. 3, Ganeshkhind, Pune 411007, India}

\begin{abstract}
We present new {\em Giant Metrewave Radio Telescope} observations at
235 MHz and 610 MHz of 18 X-ray bright galaxy groups. These
observations are part of an extended project, presented here and in 
future papers, which combines low-frequency radio and X-ray data to 
investigate the interaction between central active galactic nuclei
(AGN) and the intra-group medium (IGM). The radio images show a 
very diverse population of group-central radio sources, varying widely 
in size, power, morphology and spectral index. Comparison of the radio 
images with {\em Chandra} and {\em XMM-Newton} X-ray images shows 
that groups with significant substructure in the X-ray band and
marginal radio emission at $\gtsim$1 GHz host low-frequency radio 
structures that correlate with substructures in IGM. Radio-filled
X-ray cavities, the most evident form of AGN/IGM interaction in our
sample, are found in half of the systems, and are typically associated 
with small, low- or mid-power double radio sources. Two systems, 
NGC5044 and NGC4636, possess multiple cavities, which are
isotropically distributed around the group center, possibly due to
{\em group weather}. In other systems the radio/X-ray correlations
are less evident. However, the AGN/IGM interaction can manifest itself 
through the effects of the high-pressure medium on the morphology, 
spectral properties and evolution of the radio-emitting plasma. In
particular, the IGM can confine fading radio lobes in old/dying radio 
galaxies and prevent them from dissipating quickly. Evidence for radio 
emission produced by former outbursts that coexist with current
activity is found in six groups of the sample.
\end{abstract}


\maketitle


%
%

\begin{deluxetable*}{lccclcc}
\tablewidth{0pt}
\tablecaption{The list of galaxy groups
\label{tab:sample}
 }
\tablehead{
\colhead{Group name} & \colhead{RA$_{\rm J2000}$} &
\colhead{DEC$_{\rm J2000}$}  &  \colhead{$z$} &
\colhead{$S_{1.4 \, GHz}$ $^{a}$} & \colhead{
  log $P_{1.4 \, GHz}$} & \colhead{Scale} \\
\colhead{} & \colhead{(h m s)} & \colhead{($^{\circ}$,$^{\prime}$, $^{\prime \prime}$)}  & \colhead{} & 
\colhead{(mJy)} &
  \colhead{(W Hz$^{-1}$)} & \colhead{(kpc/$^{\prime \prime}$)} 
}
\startdata
UGC\,408  & 00 39 18.6 & +03 19 52   & 0.0147 &\phantom{0}1710 $^{b}$ & 23.92 &  0.300 \\
NGC\,315  & 00 57 48.9 & +30 21 09   & 0.0165 & \phantom{0}2010     & 24.09 &  0.336 \\
NGC\,383  & 01 07 25.0 & +21 24 45   & 0.0170 & \phantom{0}4862 $^{b}$ & 24.50 & 0.346 \\
NGC\,507 & 01 23 40.0 & +33 15 20   & 0.0165 & \phantom{000}99      & 22.78 & 0.336   \\
NGC\,741  & 01 56 21.0 & +05 37 44   & 0.0185 & \phantom{0}1066 $^{b,c}$ & 23.92 & 0.376 \\
HCG\,15   & 02 07 37.5 & +02 10 50   & 0.0228 & \phantom{000}25     & 22.47 & 0.460  \\
NGC\,1407 & 03 40 11.9 & $-$18 34 39 & 0.0059 & \phantom{000}86     & 21.82 &0.122  \\
NGC\,1587 &  04 30 39.9 & +00 39 43   & 0.0123 & \phantom{00}132     & 22.65&0.252\\
MKW\,2   &  10 30 10.7 & $-$03 09 48 & 0.0368 & \phantom{00}385     & 24.11 & 0.731 \\
NGC\,3411 &  10 50 26.1 & $-$12 50 42 & 0.0153 & \phantom{000}38     & 22.30 & 0.312 \\
NGC\,4636 &  12 42 50.4 & +02 41 24   & 0.0031 &\phantom{000}82 $^c$   & 21.24 &0.064 \\
HCG\,62  &  12 53 05.8 & $-$09 12 16 & 0.0137 &  \phantom{0000}5     & 21.32  & 0.280 \\
NGC\,5044 &  13 15 24.0 & $-$16 23 06 & 0.0090 &  \phantom{000}36     & 21.81 &0.185 \\
NGC\,5813 &  15 01 11.2 & +01 42 07   & 0.0066 &\phantom{000}15     & 21.15 & 0.135 \\ 
NGC\,5846 &  15 06 29.3 & +01 36 20   & 0.0057 & \phantom{000}21     & 21.18 & 0.118 \\
AWM\,4     &  16 04 57.0 & +23 55 14   & 0.0318 & \phantom{00}608     & 24.15 &0.624 \\
NGC\,6269 &  16 58 02.4 & +27 51 42   & 0.0348 &\phantom{000}51     & 23.15 & 0.693 \\
NGC\,7626  (NGC\,7619) $^e$ &  23 20 42.3 & +08 13 02   & 0.0114 & \phantom{00}780  & 23.36 & 0.233 \\
\\[-2mm]
\enddata
\tablecomments{
$a:$ Measured from the NVSS images, unless stated otherwise. $b:$
Condon, Cotton \& Broderick (2002). $c:$ Background radio source not
included. The flux of the unrelated source has been measured on the
FIRST image ($S_{1.4\, GHz}=39$ mJy for the source
in NGC\,741; $S_{1.4\, GHz}=14$ mJy for the source in
NGC\,4636). $d:$ Mahdavi et al. (2000). $e:$ The group is centered on
NGC\,7619 (RA=23h20m14.5s, DEC=+08$^{\circ}$12$^{\prime}$23$^{\prime\prime}$, z=0.0125).
}
\end{deluxetable*}


\section{Introduction}\label{sec:intro}

Our view and understanding of cool cores in clusters and groups of
galaxies have changed significantly with the launch of
the X-ray observatories {\em Chandra} and {\em XMM-Newton}. As
predicted by the cooling flow model (Fabian, Nulsen \& Canizares
1984; Fabian 1994), radiative cooling of the X-ray emitting hot gas indeed occurs in
the central regions of clusters, but the amount of cool gas is much less than expected 
(Peterson \& Fabian  2006 for a review), requiring a source of heat to balance the
radiative losses. Based on the detection of X-ray structures 
(such as cavities, ripples, filaments and shocks) associated with the
central radio galaxy in many cool-core systems (e.g., Fabian et al. 
2005, 2006; Mazzotta et al. 2004; Forman et al. 2005; McNamara et al. 
2005; Gitti et al. 2007; Blanton et al. 2009), radio-loud activity of the
central active galactic nucleus (AGN) is presently the leading candidate for such a source of
heat (see, for instance, the review by McNamara \& Nulsen 2007).

The combination of X-ray and radio data is vital to understand how AGN 
feedback operates. However, radio observations at frequencies of
$\sim 1$ GHz and higher often fail to detect the radio-emitting plasma presumed to
fill the X-ray cavities and produce many of the observed X-ray structures. 
Some clusters show both radio-filled and ``ghost'' cavities (for
instance, Perseus; Fabian et al. 2006), suggesting that the central galaxy has
undergone repeated radio outbursts, with the ghost cavities resulting
from buoyantly rising lobes inflated during past outburst
episodes. In
this picture, radio emission filling ghost cavities would be aged
and hence characterized by very steep spectrum, making its detection
very challenging at GHz frequencies.
Observations at frequencies $<1$ GHz indeed 
confirm the presence of relativistic plasma in the ghost cavities of a 
number of rich clusters (e.g., Fabian et al. 2002, Clarke et al. 2005 \& 2009).  
These cases emphasize the importance of multi-frequency radio observations, 
encompassing the low-frequency domain, to trace the history of AGN outbursts
and the related energy input into cluster and group cores. 

To date, most studies have focused on AGN feedback in massive 
clusters (e.g., the analysis of cluster samples by B{\^i}rzan et 
al. 2004, 2008;  Dunn \& Fabian 2004, 2006; Rafferty et al. 2006; 
Diehl et al. 2008; Mittal et al. 2009), but
the majority of galaxies in the Universe reside in smaller
units, such as poor clusters and groups (Eke et al. 2004). 
As in clusters, X-ray bright groups are often dominated by
giant elliptical or cD galaxies which host AGNs. 
As the group gravitational potential is shallower, the impact of AGN outbursts
can be severe, with relatively small energy injections causing
dramatic effects on the energetics and spatial distribution of the 
intra-group medium possibly even ejecting
gas from the group (Giodini et al. 2010, Lanz et al. 2010). 
Groups are therefore a key environment to assess the influence of 
AGN heating on the thermal history of galaxies and their surrounding 
gaseous media. 

With the aim of extending the investigation of the hot gas/AGN interplay 
to the group environment and low-frequency radio regime, we undertook
a study of a sample of 18 groups that combines new low-frequency {\em Giant
Metrewave Radio Telescope} ({\em GMRT}) observations and X-ray data
from {\em Chandra}  and {\em XMM-Newton} observations. Our project will be
presented here and in future papers (O'Sullivan et al. 2011,
Giacintucci et al. and O'Sullivan et al. in preparation). In this paper, we present the
{\em GMRT} observations at 235 MHz and 610 MHz for our sample, 
showing the new radio images and providing basic radio properties of
the central radio sources. 
We also compare the radio images with simple X-ray intensity images to
demonstrate correlations, when found, with the intra-group medium
(IGM) substructures, which previous studies have shown to be associated 
with heating and gas motions (X-ray cavities, filaments, ripples, fronts).
In future papers, we will analyze these correlations in more detail, examine 
the spectral properties, pressure and ages of the radio sources, and
consider their impact on the groups in which they reside, and in
particular on the properties of the IGM.

The present paper is organized as follows: in Section \ref{sec:sample} we present 
the sample of galaxy groups; the {\em GMRT} observations 
and data reduction are described in Section \ref{sec:obs}; Section 4
summarizes the X-ray ({\em Chandra} and {\em XMM-Newton}) data reduction;
the radio images and the radio/X-ray comparison are 
presented in Section \ref{sec:images}; 
the discussion and summary are given in Sections 
\ref{sec:disc} and \ref{sec:summary}, respectively. 
Throughout the paper we assume H$_0$ = 70 km sec$^{-1}$ Mpc$^{-1}$, 
$\Omega_m$ = 0.3, and $\Omega_{\Lambda}$ = 0.7. The radio spectral index 
$\alpha$ is defined according to $S{_\nu}\propto \nu^{-\alpha}$, where
$S_{\nu}$ is the flux density at the frequency $\nu$.



\begin{deluxetable*}{lcccccc}
\tablewidth{0pt}
\tablecaption{Details of the GMRT observations
\label{tab:obs}
 }
\tablehead{
\colhead{Group name} & \colhead{Observation} &
\colhead{Frequency}  &  \colhead{Bandwidth} & \colhead{Integration} &
\colhead{Beam, PA $^a$} & \colhead{rms}  \\
\colhead{} & \colhead{date} & \colhead{(MHz)}  & \colhead{(MHz)} &
\colhead{time (min)}& \colhead{ ($^{\prime \prime} \times^{\prime \prime}$, $^{\circ}$)} & \colhead{(mJy beam$^{-1}$)} 
}
\startdata
UGC\,408 & Aug 2007 & 610 & 32 & 110 & 6.4$\times$5.2, 56 & 0.10  \\
         & Aug 2008 & 235 & \phantom{0}8  & 120 & 15.4$\times$12.5, 26 & 0.40 \\

NGC\,315 & Feb 2008 & \phantom{0}610 $^b$ & 32 & 380 & 5.2$\times$5.0, 61 & 0.10  \\
         & Aug 2008 & \phantom{0}235$^b$ & \phantom{0}8  & 280 & 15.0$\times$15.0, 0 & 0.70  \\

NGC\,383 & Feb 2008 & 610 & 32 & 180 & 4.7$\times$3.9, 53 & 0.12  \\
         & Aug 2009 & 235 &  \phantom{0}8 &160 & 15.0$\times$11.9,$-$74 & 1.00 \\

NGC\,507 & Jul 2006 & 610 & 32 & 220 & 7.3$\times$5.7, 62  & 0.05 \\
         & Aug 2008 & 235 &  \phantom{0}8 & 120 & 17.7$\times$14.5, 62  & 1.00 \\

NGC\,741 & Aug 2006 & 610 & \phantom{0}32 $^c$ & 140 & 7.9$\times$4.6, 52 & 0.05  \\
         & Aug 2007 & 235 & \phantom{0}8  & 140 & 12.7$\times$12.3, 64 & 0.30  \\

HCG\,15  & Aug 2006 & 610 & \phantom{0}32 $^c$ &240 & 8.0$\times$4.7, 63  &  0.04  \\
         & Aug 2008 & 235 & \phantom{0}8       & 270 &  14.2$\times$11.5, 61 &  0.33 \\

NGC\,1407 & Jul 2006 & 610 & \phantom{0}32 $^d$ & 200 & 5.6$\times$4.3, 41 & 0.10  \\
         & Aug 2008 & 235 &  \phantom{0}8      &  130 & 15.4$\times$12.5, 15 & 0.50 \\

NGC\,1587 & Aug 2006 & 610 & \phantom{0}32 $^c$ & 200 & 5.7$\times$4.7, 67 & 0.05 \\
         & Aug 2008 & 235 &  \phantom{0}8      & 120 & 17.2$\times$11.0, 46 & 1.00 \\

MKW\,2 $^e$  & Aug 2003 & 610 & 16      & 100 & 5.1$\times$4.6, 42 & 0.13 \\
         & Jul 2005 & 235 & \phantom{0}8 & 100 & 12.1$\times$9.7, 65 & 0.65 \\

NGC\,3411 & Aug 2006 & 610 & 32 & 80 & 6.9$\times$5.3, $-$19  & 0.09 \\
         & Feb 2008 & 235 & \phantom{0}8  & 160 & 14.0$\times$13.0, 0 & 0.40  \\

NGC\,4636 $^f$& Aug 2006 & 610 & 32 & 120 & 5.8$\times$4.3, 48 & 0.05  \\
  & Feb 2008 & 235 & \phantom{0}8  & 100 & 12.0$\times$11.5, 14 & 0.20 \\

HCG\,62 $^g$ & Feb 2008 & 610 & \phantom{0}32 $^c$ & 100 & 6.9$\times$ 5.4, $-$39 & 0.05 \\
 & Feb 2008 & 235 & \phantom{00}8  &100 & 14.1$\times$12.3, 47    & 0.17  \\

NGC\,5044 $^h$& Feb 2008 & 610 & \phantom{0}16 $^i$  & 130 &  16.8$\times$11.7, $-$4 & 0.05\\
        & Feb 2008 & 235 & \phantom{00}8 $^i$  &  140 &  13.7$\times$11.5, 10  & 0.25 \\

NGC\,5813 $^j$&  Aug 2008  & 235 &   \phantom{0}8  & 100  &  16.5$\times$15.1, 74  &  0.30\\

NGC\,5846 & Aug 2006 & 610 & \phantom{0}32 $^c$ & 140 & 14.8$\times$6.1, 56 &  0.04  \\

AWM\,4 $^k$  &  Aug 2006 & 610 &  32 & 160  &  5.0$\times$4.0, 43 & 0.05  \\
         &  Jun 2006 & 327 & 32 & 100   &  9.0$\times$7.8, 53 & 0.40 \\
         &  Jul 2006 & 235 & 16 & 120 & 12.7$\times$10.4, 75  & 0.80 \\

NGC\,6269 & Feb 2008 & 610 & \phantom{0}16 $^i$ & 260 & 5.3$\times$4.1, 69 & 0.07  \\
         & Feb 2008 & 235 &  \phantom{00}8 $^i$ & 260 & 14.1$\times$12.0, 80 & 0.60 \\
NGC\,7626& Aug 2007 & 610 & 32 & 100 & 6.2$\times$4.9, 31  & 0.05 \\
         & Aug 2008 & 235 &  \phantom{0}8 & 120 & 14.2$\times$12.0, 57
         & 0.80 \\
\enddata
\tablecomments{$a:$ FWHM. $b:$ Observed with two pointings (see text for details).
$c,d: $ The observations were performed using a total bandwidth of 32 MHz 
(USB+LSB), but only the USB ($c$) and LSB ($d$) data set was used for the analysis. 
$e:$ Some results based on these observations have been presented in Giacintucci et al. (2007).
$f:$ Some results based on the 610 MHz observations have been
presented in Baldi et al. (2009a).
$g:$  The observations of this object have been presented in
Gitti et al. (2010).
$h:$  The observations of this object have been presented in
David et al. (2009, 2011).
$i:$ Observed in dual 235/610 MHz mode.
$j:$ Some results based on these observations have been presented in
Randall et al. (2011).
$k:$ The observations of this object have been presented in
Giacintucci et al. (2008) and O'Sullivan et al. (2010a, 2010b).
}
\end{deluxetable*}



\begin{deluxetable}{lccc}
\tablewidth{0pt}
\tablecaption{List of the X-ray observations used for the images.
\label{tab:obs}
 }
\tablehead{
\colhead{Group name} & \colhead{Instrument} & \colhead{Observation ID} &
\colhead{Exposure time}  \\
\colhead{} & \colhead{} & \colhead{}  & \colhead{(ksec)} \\
}
\startdata
UGC\,408 & {\em Chandra ACIS-S} & 4053 & 29 \\
NGC\,315 & {\em XMM-Newton } & 0305290201 & 52 \\
NGC\,383 & {\em XMM-Newton}    & 0305290101  & 24 \\
NGC\,507 & {\em Chandra ACIS-I} & 2882 & 44 \\
NGC\,741 & {\em Chandra ACIS-S} & 2223 & 31 \\
                            &  {\em XMM-Newton}    &0153030701 & 9 \\
HCG\,15   & {\em XMM-Newton}  & 0052140301	& 35 \\
NGC\,1407 & {\em Chandra ACIS-S} & 791 & 49 \\
NGC\,1587 & {\em Chandra ACIS-I} & 2217 & 20\\
MKW\,2 & {\em XMM-Newton}  & 0404840201 & 48 \\
NGC\,3411 &  {\em Chandra ACIS-S} & 3243 & 30 \\
NGC\,4636 &  {\em Chandra ACIS-I} & 4415 & 75 \\
HCG\,62 & {\em Chandra ACIS-S} & 921 & 49 \\
NGC\,5044 & {\em Chandra ACIS-S} & 9399 & 84 \\
NGC\,5813 &  {\em Chandra ACIS-S} & 9517 & 100 \\
NGC\,5846 & {\em Chandra ACIS-I} & 7923 & 91 \\
NGC\,6269 & {\em Chandra ACIS-I} & 4972 & 40 \\
NGC\,7626 & {\em Chandra ACIS-I} & 2074 & 27 \\
\enddata
\end{deluxetable}



\begin{deluxetable}{lccc}
\tablewidth{0pt}
\tablecaption{Radio source data
\label{tab:flux2}
 }
\tablehead{
\colhead{Source} & \colhead{$S_{\rm 235\,MHz} \pm 8\%$} &
\colhead{$S_{\rm 610\,MHz} \pm5\%$}  &  \colhead{$\alpha_{\rm
    235\,MHz}^{\rm 610\,MHz}$} \\
\colhead{} & \colhead{(mJy)} & \colhead{(mJy)}  &
\colhead{($\pm$0.10)} \\
}
\startdata
UGC\,408  & \phantom{0}5260 & \phantom{0}3184 &
\phantom{0}0.53 \\ 
NGC\,315 & 15411 & $>$2500 $^a$&  $< 1.91 ^b$ \\ 
NGC\,383 &  17815 &  \phantom{0}8206  &  \phantom{0}0.81 \\
NGC\,507 &  \phantom{0}1100 & \phantom{00}372 &
\phantom{0}1.14  \\
NGC\,741 & \phantom{00}5734 $^c$ & \phantom{00}2192 $^c$ & \phantom{0}1.01 \\
HCG\,15  & \phantom{00}$-$  &  \phantom{0000}50 $^{d}$  & $-$ \\
NGC\,1407 & \phantom{0}1490 & \phantom{00}150 & \phantom{000}2.41 $^e$ \\
NGC\,1587 & \phantom{00}655  & \phantom{000}222
$^d$& $> 1.13$ $^f$\\
MKW\,2   & \phantom{0}1951 & \phantom{00}675 & \phantom{0}1.11 \\
NGC\,3411 & \phantom{00}555 & \phantom{00}140 & \phantom{0}1.44 \\
NGC\,4636 & \phantom{000}248 $^c$& \phantom{000}139 $^c$ & \phantom{0}0.61 \\
HCG\,62 &  \phantom{000}42  & \phantom{000}14 & \phantom{0}1.15  \\
NGC\,5044 & \phantom{00}229 & \phantom{000}38 & \phantom{000}1.88 $^e$\\
NGC\,5813 & \phantom{000}81  & \phantom{0000}$-$ & $-$\\
NGC\,5846 & \phantom{0000}$-$ & \phantom{000}36 & $-$\\
AWM\,4    & \phantom{0}2750 & \phantom{0}1450 &
\phantom{0}0.67 \\
NGC\,6269 & \phantom{00}230 & \phantom{00}115 & \phantom{0}0.73  \\
NGC\,7626 & \phantom{00}3444 $^c$& \phantom{00}1597 $^c$ & \phantom{0}0.81 \\
\enddata
\tablecomments{
$a:$ The flux density measured on the 610 MHz is underestimated.
$b:$ Using the {\em WSRT} 609 MHz flux by Mack et al. (1998), $S_{\rm
  609\,MHz} = 5.3$ Jy,  the source has $\alpha=1.11$ in the 235-609 MHz range.
$c:$ Background radio source subtracted.
$d:$ Measured on the low-resolution image.
$e:$ This value has to be considered an overestimate of the real spectral
index of the source because part of the extended emission visible at
235 MHz is not detected at the sensitivity level of the 610 MHz image.
$f:$ The 235 MHz observation detects only the central point source.}
\end{deluxetable}


\begin{deluxetable*}{lcccccc}
\tablewidth{0pt}
\tablecaption{Radio and X-ray properties of the groups
}
\tablehead{
\colhead{Source} & \colhead{ log $P_{\rm 235 \,  MHz}$} &
\colhead{LLS $^{a}$} & \colhead{Radio}  & \colhead{log $L_X$}&
\colhead{X-ray cavities $^{b}$}& \colhead{Class $^{c}$}\\ 
\colhead{} & \colhead{(W Hz$^{-1}$)} &\colhead{(kpc)} &
\colhead{morphology} & \colhead{(erg
  s$^{-1}$)}  & \colhead{}  & \\
}
\startdata
UGC\,408  & 24.41  & \phantom{00}80 & double + diffuse cocoon
&  \phantom{0}41.40 $^{d}$& $-$ & 1\\
NGC\,315 & 24.98 & 1160 & giant double with distorted tails  &
41.57 &
$-$ & 2\\
NGC\,383 & 25.06 & \phantom{0}900 &  giant double with
distorted tails& 42.72 & $-$ & 2\\
NGC\,507 & 23.83 & \phantom{00}70 & asymmetric double & 42.95&
1 (filled) $^{e}$ & 1\\
NGC\,741 & 24.65 & \phantom{0}210 & complex & 42.14 & 1
(ghost) $^{f}$ & 1, 2\\
HCG\,15  & $-$ & \phantom{000}150 $^{g}$ & point source +
diffuse emission & 41.72 & $-$ & 3\\
NGC\,1407 & 23.06 & \phantom{00}80 & double + diffuse emission
& 41.23 & 1 (filled) $^{h}$ & 1, 3\\ 
NGC\,1587 &23.35 & \phantom{0000}22 $^{g}$& point source +
diffuse emission & 40.92 & $-$ & 3\\
MKW\,2   & 24.79 & \phantom{0}670 & lobe-dominated double &
42.32 &$-$ & 2\\
NGC\,3411 & 23.47 & \phantom{00}80  & point source +
diffuse emission  & \phantom{0}42.45 $^{i}$ & $-$ & 3\\
NGC\,4636 & 21.72 & \phantom{00}10 & asymmetric double & 41.88 &
multiple (filled and ghost) $^{j}$ & 1 \\
HCG\,62 &  22.25 & \phantom{00}33 & asymmetric double & 42.69
& 2 (filled) $^{k}$ & 1 \\
NGC\,5044 & 22.62 & \phantom{00}63 & complex & 42.81& multiple
(filled and ghost) $^{l}$ & 1\\
NGC\,5813 & 21.90 & \phantom{00}22 & double-double  & \phantom{0}42.00
$^{m}$& multiple
(filled and ghost) $^{n}$  & 1\\
NGC\,5846 & $-$ &
\phantom{000}12 $^o$& point source + extension & 41.71 & 2 (filled)
$^{p}$ + 1 (ghost)$^{q}$ & 1 \\
AWM\,4    & 24.80 &  \phantom{0}160 & WAT &  \phantom{0}43.64
$^{r}$ & 1 (filled) $^{s}$ & 1,2 \\
NGC\,6269 & 23.81   & \phantom{00}36 & double& 43.20 & 2
(filled) $^t$ & 1\\
NGC\,7626 & 24.00 & \phantom{0}185  & double with distorted
tails &42.05 & $-$ & 2 \\
\enddata
\tablecomments{
$a:$ Largest linear size (LLS), measured on the 235 MHz image unless indicated otherwise.
$b:$ Number and type of X-ray cavities as reported in the literature.
$c:$ As defined in Sect.~6.4.
$d:$ Diehl \& Statler (2007). 
$e:$ A possible cavity has been identified by Dong, Rasmussen \&
Mulchaey (2010) at the base of the west lobe.
$f:$ Jetha et al. (2008).
$g:$ Measured on the low-resolution image at 610 MHz.
$h:$ A possible small cavity has been identified by Dong, Rasmussen \&
Mulchaey (2010) at only $\sim$0.3 kpc from the center.
$i:$ Mahdavi et al. (2000).
$j:$  Baldi et al. (2009a).
$k:$ Gitti et al. (2010).
$l:$ David et al. (2009).
$m:$ Popesso et al. (2004).
$n:$ Randall et al. (2011).
$o:$ Measured on the low-resolution image.
$p:$ Dong, Rasmussen \&
Mulchaey (2010).
$q:$ Machacek et al. (2011).
$r:$  Ebeling et al. (1998).
$s: $ O'Sullivan et al. (2010a).
$t:$ Baldi et al. (2009b); the radio lobes are coincident with depressions in the X-ray
brigthness, but the {\em Chandra} data are not deep enough to confirm
that these are X-ray cavities.}
\end{deluxetable*}

\section{The sample of galaxy groups}\label{sec:sample}

Using our own and archival {\em Chandra} and {\em XMM-Newton} observations, in 
combination with literature and archival radio images, we have
selected 18 nearby ($z<0.05$), elliptical-dominated groups as targets for our joint {\em
  GMRT}/X-ray study. The groups were chosen to possess structures, either in the X-ray
brightness and temperature distribution or radio morphology,
which strongly indicate ongoing or past AGN activity and thus
interaction between the radio source 
and surrounding IGM. We emphasize that this is not a statistical
sample -- our selection of systems, in which the short-lived disturbed
features possibly associated with AGN feedback are still apparent,
prevents this -- but is instead 
designed to provide as thorough a picture as possible of the variety
of forms of AGN/IGM interaction taking place in groups. The source list is
presented in Table 1, where, unless stated otherwise, the 1.4 GHz flux
density is measured from the NRAO {\em VLA} Sky Survey (NVSS; Condon et al. 1998) images. 

The sample includes a number of well-known radio sources, such as 3C\,31 (NGC\,383) and 
B2~0055+30 (NGC\,315), for which pointed, multi-frequency and
multi-resolution radio observations are available in the 
literature, usually at frequencies $\ge$1 GHz. All groups in Table 1 have X-ray 
data of good quality and show temperatures in the $\sim$1-3 keV range, 
as expected for poor clusters and groups of galaxies. The velocity
dispersions of the groups are in the range $\sim$100-600 km s$^{-1}$.

\section{{\em GMRT} radio observations and data reduction}
\label{sec:obs}

The groups listed in Table 1 were observed with the {\em GMRT} at 235 MHz and 610
MHz. The observation details are summarized in Table 2, which reports source name,
observing date, frequency, total frequency bandwidth, integration time 
on source, beam (synthesized full-width half-maximum, FWHM) of 
the full array, and rms level (1$\sigma$) at full resolution. 
The poor cluster MKW\,2 was observed by Giacintucci et al. (2007) as part of a 
project devoted to the study of cD radio galaxies in rich and poor clusters. 
The radio data for AWM\,4 have been presented and analyzed in Giacintucci et 
al. (2008) and O'Sullivan et al. (2010a, 2010b). The {\em GMRT} images
of NGC\,5044, HCG\,62 and NGC\,5813 have been presented in David et
al. (2009, 2011), Gitti et al. (2010) and Randall et al. (2011),
respectively. Some results based on the 610 MHz observations of
NGC\,4636 have been presented in Baldi et al. (2009a).

The data at 610 MHz were recorded using both the upper and lower side 
bands (USB and LSB), providing a total observing bandwidth of 32 MHz. A single 
band of 8 MHz width was used for the observations at 235 MHz. NGC\,5044 and NGC\,6269 
were observed using the {\em GMRT} in dual 235/610 MHz mode, with a 16 MHz-band at 610 MHz 
and a 8 MHz-band at  235 MHz. The data at all frequencies were collected using the default 
spectral-line mode, with 128 channels for each band, resulting in a spectral resolution of 125 
kHz/channel at 610 MHz and 62.5 kHz/channel at 235 MHz. The data reduction and analysis
was carried out using the NRAO Astronomical Image Processing System (AIPS) package.  
After an initial editing of the data using SPFLG in AIPS to identify and remove bad channels 
and data affected by radio frequency interference (RFI), the data were calibrated. The flux density scale 
was set using amplitude calibrators, observed at the beginning and end of each observing run, and the 
scale of Baars et al. (1977). The bandpass calibration 
was obtained using the flux density calibrators. A central channel free of RFI was used to
normalize the bandpass for each antenna. Residual RFI  affecting the data at 235 MHz was 
removed using the task FLGIT in AIPS and subsequent accurate editing of the filtered 
data. 

After bandpass calibration, the central 84 channels were averaged to 6 channels 
of $\sim$2 MHz each at 610 MHz, and $\sim$0.9 MHz at 235 MHz, to 
reduce the size of the data set, and at the same time to minimize the bandwidth 
smearing effects within the primary beam of the {\em GMRT} antenna. After further careful 
editing in the averaged data sets, a number of phase-only self--calibration cycles 
and imaging were carried out for each data set. The large field of view of the 
{\em GMRT} required the implementation of multi-field imaging in each step of the data reduction, 
using 25 facets covering a field of $\sim2.7^{\circ} \times 2.7^{\circ}$ at 235 MHz and
$\sim 1.4^{\circ} \times 1.4^{\circ}$ at 610 MHz. 
The USB and LSB were calibrated separately. The final data sets were 
further averaged from 6 channels to 1 single channel\footnote{The bandwidth smearing, relevant 
only at the outskirts of the wide field, does not significantly affect the region 
occupied by the central radio source.}, and then combined together to produce the final images for each
group. Given the large angular sizes of NGC\,315, NGC\,383, and
MKW\,2,
the USB and LSB images were produced using the 6 channel data sets, and the final images 
were obtained by combination of the USB and LSB in the plane of the image using the task 
LTESS in AIPS. The final images of all groups were corrected for the primary beam pattern of the
{\em GMRT} antennas.

The full resolution of the {\em GMRT} is $\sim 6^{\prime \prime}$ at 610 MHz
and $\sim 13^{\prime \prime}$ at 235 MHz. The u-v range of our observations 
($\sim$0.05--20 k$\lambda$ and $\sim$0.1--50 k$\lambda$ at 235 and 
610 MHz, respectively) ensures the detection of structures with angular
size $\ltsim$ 44$^{\prime}$ (235 MHz) and $\ltsim$ 17$^{\prime}$ (610 MHz).
Beyond the full resolution images, produced with uniform weighting and
no tapering, for each source we obtained images with different resolutions,
tapering the u-v data by means of the parameters {\tt robust} and 
{\tt uvtaper} in the task IMAGR.  The rms noise level (1$\sigma$) achieved in the final full resolution images is in 
the range 35-130 $\mu$Jy at 610 MHz, and 0.2-1 mJy at 235 MHz (Table 2). 
The noise in the full resolution and lower resolution images is comparable in most
cases. The spread in the noise level depends mostly on the total time on source, residual RFI 
still affecting the data, usable bandwidth (in a number of cases only one of
the two bands provided good data; see notes in Table 2), and presence of bright sources in the field limiting the 
achievable dynamic range. 

The average residual amplitude errors are $\le$ 5\% at 610 MHz and
$\le$8\% at 235 MHz (e.g., Chandra et al. 2004). Therefore we can
conservatively assume that the absolute flux density
calibration is within 5$\%$ and 8$\%$ at 610 MHz and 235 MHz,
respectively.


\section{X-ray data reduction}

The X-ray observations, their analysis and 
more detailed interpretation will be 
described in a future paper (O'Sullivan et al. in preparation). 
We therefore provide here only a short general description of the
procedure used to create the images. 
Since our goal is to compare the radio and X-ray structures, we choose
smoothing scales, color scales, energy band, and whether or not to perform
point-source subtraction, on the basis of the sensitivity of the resulting
image.

Table 3 lists the general properties of the observations used to
produce the X-ray images shown in Section~\ref{sec:images}.
\textit{Chandra} data were reduced following methods similar to those
described in O'Sullivan et al. (2010a) and the \textit{Chandra}
analysis threads\footnote{http://asc.harvard.edu/ciao/threads/index.html}. 
A summary of the \textit{Chandra} mission and instrumentation can be found in
Weisskopf et al. (2002). 
The level 1 events files were reprocessed, bad
pixels and events with \textit{ASCA} grades 1, 5 and 7 were removed, and the
cosmic ray afterglow correction was applied. Very faint mode background filtering was
applied as appropriate. The data were corrected by the appropriate gain
map, the standard time-dependent gain and charge-transfer inefficiency
(CTI) corrections were made, and a background light curve was produced.
Periods when the background deviated from the mean by more than 3$\sigma$
were excluded. Point sources were identified using the \textsc{ciao} task
\textsc{wavdetect} from the 0.3-7.0~keV images and monoenergetic exposure maps
with energies chosen to match the mean photon energy of the data (typically
$\sim$1~keV) were produced. Detection thresholds were set to produce $\leq$1 false source
on either the ACIS-I array, or S3 CCD, depending on which was in use. Soft
band (0.3-2.0 keV) images were extracted, point sources were generally excluded (excepting
any point source thought to be related to the AGN) and the resulting holes
filled using the \textsc{dmfilth} task. The images were finally corrected
using the monoenergetic exposure maps.

In one case, UGC~408, the available \textit{Chandra} observation (ObsID 4053) 
suffers from severe background flaring, such that the
exclusion of periods of high background resulted in a poor quality image
containing few counts. We therefore chose to extract the image from the
full dataset, including the flare periods.

\textit{XMM-Newton} data were reduced following methods similar to those
described in O'Sullivan et al. (2007).
A detailed summary of the \textit{XMM-Newton} mission and instrumentation
can be found in Jansen et al. (2001, and references therein). 
The raw data
from the EPIC instruments were processed with the \textit{XMM} Science
Analysis System (SAS) tasks \textsc{epchain} and \textsc{emchain}. Bad
pixels and columns were identified and removed and the events lists
filtered to include only those events with FLAG = 0 and patterns 0-12 (for
the MOS cameras) or 0-4 (for the PN). Periods when the background count
rate deviated from the mean by more than 3$\sigma$ were excluded. Soft
band (0.3-2.0 keV) images and monoenergetic exposure maps (with energies chosen as for
\textit{Chandra} data) were extracted. Images of the particle component of
the background, determined from the ``telescope closed'' datasets of Marty et
al. (2003), were extracted and scaled to match the count rate in areas of
the detector outside the telescope field of view. Point sources were
identified using the \textsc{edetect\_chain} script and, where desirable,
excluded from further analysis using circular regions of radius
25$^{\prime\prime}$.  Where out-of-time (OOT) events produced a significant
readout trail in the EPIC-pn camera, an OOT events list was created using
\textsc{epchain}, and appropriately scaled images were used to
statistically subtract the trail. Where smoothed images were required, the
soft band images were adaptively smoothed using the \textsc{asmooth} task,
with signal-to-noise ratios of 10-20, and the resulting smoothing scales
applied to particle-subtracted, OOT-subtracted images. Finally, unsmoothed
images were corrected using the monoenergetic exposure maps.


\section{The GMRT radio images and radio/X-ray comparison}\label{sec:images}

In this section we present the new {\em GMRT} images at 235 MHz and 610 MHz
for all groups listed in Table 1, with the exception of AWM4, published in
Giacintucci et al. (2008). The radio images are overlaid on the red-band optical images from the
second Palomar Observatory Sky Survey (POSS-2) and on the X-ray images
from {\em Chandra} or {\em XMM-Newton} observations. We also mention the conversion factor
from angular to linear scale (Table 1) in the caption of each figure.

In Table 4, we provide the 235 MHz and 610 MHz total flux densities,
and spectral index $\alpha$ between these two frequencies for each
radio source. The flux densities were measured on primary-beam corrected
images with similar resolution at both frequencies. Given the high
signal-to-noise ratio in all our images, the error associated with 
the flux density measurement is dominated by the uncertainty in the residual amplitude 
calibration errors (5\% at 610 MHz and 8\% at 235 MHz;
Sect.~\ref{sec:obs}).

Table 5 summarizes the most relevant observational properties of the groups
in the radio and X-ray bands, i.e., radio power at 235 MHz, radio
largest linear size (LLS), radio morphology,  group X-ray luminosity
($L_X$), presence and type of X-ray cavities as reported in the literature\footnote{the search and
 identification of possible X-ray cavities not reported in the
literature are deferred to a future paper (O'Sullivan et al. in preparation).}, and class as defined in Sect.~6.4.
Unless indicated otherwise in the caption of the table, the LLS was measured on the
235 MHz image, and the X-ray luminosity of the corresponding
group is from Mulchaey et al. (2003).

\subsection{UGC\,408}\label{sec:ugc408}

UGC\,408 (also known as NGC\,193) is a central dominant early-type
galaxy, classified as a lenticular in Third Reference Catalogue of
Bright Galaxies (RC3; de Vaucouleurs et al. 1991) of a poor group of
galaxies, HDCE 25.
It hosts the Fanaroff-Riley type I (FR I; Fanaroff \& Riley 1974) 
radio source 4C\,+03.01. 

\begin{figure*}
\centering
\epsscale{1}
\plotone{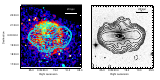}
\caption{UGC\,408. {\em Left:} {\em GMRT} 610 MHz full-resolution contours
(FWHM=$6.4^{\prime\prime} \times 5.2^{\prime\prime}$, p.a. $56^{\circ}$; 
1$\sigma$=100 $\mu$Jy beam$^{-1}$), overlaid on the smoothed, 0.3-2.0 keV {\em Chandra} 
image. Radio contours are spaced by a factor 2, starting from +3$\sigma$. 
{\em Right:} {\em GMRT} 235 MHz low-resolution contours (FWHM=$24.0^{\prime\prime} \times 21.0^{\prime\prime}$, 
p.a. $0^{\circ}$; 1$\sigma$=500 $\mu$Jy beam$^{-1}$) on the {\em POSS-2} optical image. 
Radio contours (black and white) are spaced by a factor 2, starting
from $+3\sigma$. The $-3\sigma$ level is shown as dashed black contours. For this source 
the scale is 0.300 kpc/$^{\prime\prime}$.}
\label{fig:ugc408}
\end{figure*}

The {\em GMRT} 610 MHz full resolution contours are
shown in Fig.~\ref{fig:ugc408} (left), overlaid on the smoothed {\em
  Chandra} image. In the right panel we show the 235 MHz image at the 
resolution of $24^{\prime \prime} \times 21^{\prime \prime}$, overlaid 
on the optical image. The radio source has two bright and straight jets with a 
total extent of $\sim$ 80 kpc (side to side). 
The eastern jet appears significantly 
brighter than the western one. This asymmetry is also observed at 1.4 GHz with the 
{\em Very Large Array} ({\em VLA}), and on the parsec scale with the
{\em Very Long Baseline Array} ({\em VLBA}) 
at 1.7 GHz (Xu et al. 2000), where 4C\,+03.01 exhibits a core-jet morphology aligned with the 
large scale structure. The two jets are embedded in a low-surface brightness ``cocoon'' that extends 
perpendicular to the jet axis, out to a projected distance of $\sim$30 kpc from 
the center, as measured on the 235 MHz image. The sharp edges at
  the end of the two jets suggest that they are impinging on the
  external medium.

The {\em Chandra} image in Fig.~\ref{fig:ugc408} (left) shows a $\sim 20$
kpc-radius bright ring around the central X-ray nucleus. This feature
may be the outer shell of a large single X-ray cavity projected at the group
center, although it may also be the result of the superposition of two
cavities along the line of sight. The radio jets extend beyond the
possible central cavity and show little, if any, correlation with
the substructures detected in the X-ray image.  On the contrary, a
correlation is observed between the outer border of the cocoon and the
bright X-ray rim of the candidate cavity in the northern region. Although less
evident, a similar correlation is also present in the southern and
eastern part. This suggests that the cavity may be filled by the radio
plasma in the diffuse cocoon with the X-ray bright rim representing
an edge-brightened shell of gas.

\subsection{NGC\,315}\label{sec:ngc315}

NGC\, 315 is a giant elliptical at the core of the poor group
WBL~22 in the Group Evolution Multiwavelength Study (GEMS) sample 
(Forbes et al. 2006), which is part of the 
Pisces-Perseus supercluster.
This galaxy is the optical counterpart
of the well-known giant FR I radio galaxy B2\,0055+30, which has been studied 
at many frequencies and angular resolutions (e.g., Bridle 1976, 1979; 
Willis et al. 1981; Venturi et al. 1993; Mack et al. 1997, 1998; Xu et al. 
2000; Laing et al. 2006; Worrall et al. 2007). 

Given the large angular size of B2\,0055+30, the
{\em GMRT} observations were carried out in mosaic mode
to obtain uniform sensitivity, covering 
the region of interest with two fields centered on RA=00h57m30.0s, 
DEC=$+30^{\circ} \, 24^{\prime} \, 00^{\prime \prime}$ and RA=00h59m00.0s, 
DEC=$+30^{\circ} \,10^{\prime} \, 00^{\prime \prime}$. Each field was calibrated 
separately and the final images were produced with the same restoring beam, 
primary-beam corrected and then combined in the plane of the image.

\begin{figure*}
\centering
\epsscale{1}
\plotone{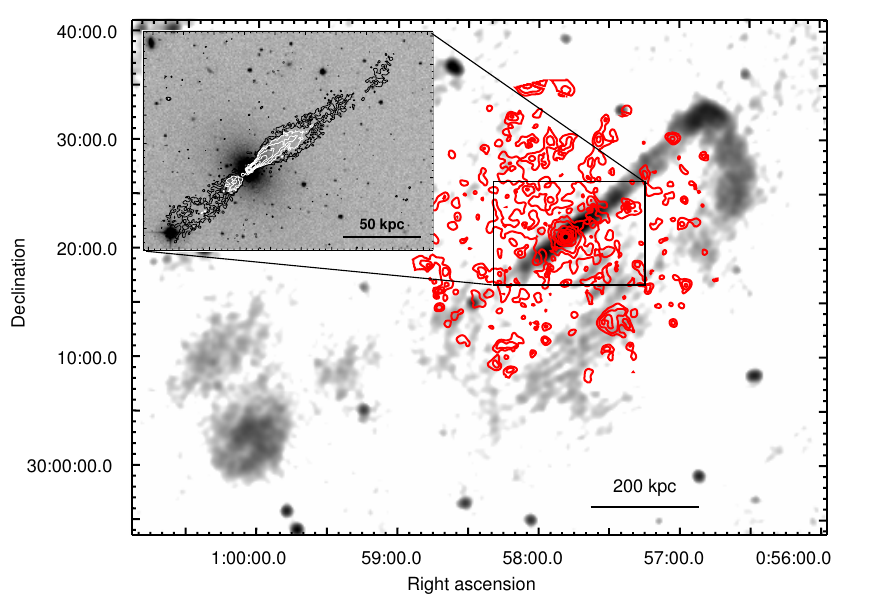}
\caption{NGC\,315. {\em GMRT} 235 MHz low-resolution, gray-scale image (FWHM= 
$40.0^{\prime\prime} \times 40.0^{\prime\prime}$, p.a. $0^{\circ}$; 
1$\sigma$=1.5 mJy beam$^{-1}$). The smoothed, 0.3-2.0 keV 
{\em XMM-Newton} image (with point sources removed) is shown as red
contours starting at 3$\sigma$ above the background and increasing by
a factor of 2. The inset shows the {\em GMRT} 610 MHz 
full-resolution contours (FWHM=$5.2^{\prime\prime} \times 
5.0^{\prime\prime}$, p.a. $61^{\circ}$; 1$\sigma$=100 $\mu$Jy beam$^{-1}$)
of the innermost region, overlaid on the POSS-2 optical image. 
Radio contours are spaced by a factor 2, starting from 0.4 mJy beam$^{-1}$. 
For this source the scale is 0.336 kpc/$^{\prime\prime}$.}
\label{fig:ngc315}
\end{figure*}

Fig.~\ref{fig:ngc315} shows the large-scale radio emission of NGC\,315
as deteced in our low-resolution (FWHM=40$^{\prime \prime}$) mosaic at 
235 MHz  (gray-scale image). The inset presents the full resolution contours at 610 MHz, overlaid 
on the optical image, showing the radio core and the inner jets. As observed in 
similar resolution images
(e.g., Mack et al. 1997), at 235 MHz the north-west jet propagates straight and continuous 
to a bright hot-spot at a distance of $\sim$360 kpc ($\sim 18^{\prime}$) from the galaxy nucleus.
The south-east jet is much fainter and appears intermittently. The two
lobes are also very asymmetric, with the north-west one 
brighter, narrower, and more prominently distorted. The lobe is heavily curved 
in a ``back-flow'' direction and forms a low-brightness tail extending parallel 
to the jet. The total linear size of the source is $\sim$ 1.2 Mpc (side to side), 
however it reaches the exceptionally large extent of $\sim$ 2 Mpc, if we 
also include the length of the radio tails beyond the two sharp bends. 

The radio properties are summarized in Table 4. A total flux of 2.5 Jy
was obtained at 610 MHz from an image (not shown here) produced 
with a similar resolution as the 235 MHz image in Fig.~\ref{fig:ngc315}. Since 
the south-east lobe is not detected at 610 MHz, our 610 MHz flux density is
underestimated.  By comparison, Mack et al. (1997) report a flux of
5.3 Jy at 609 MHz with the {\em Westerbork Synthesis Radio Telescope} ({\em WSRT}).

The X-ray emission of NGC\,315 was analyzed at high resolution by Worrall, 
Birkinshaw \& Hardcastle (2003) and Worrall et al. (2007) using {\em Chandra} data, 
which revealed a prominent X-ray jet coincident with the first $\sim30^{\prime \prime}$ of the 
brighter radio jet.  A smoothed {\em XMM-Newton} image is shown as
red contours in Fig.~\ref{fig:ngc315}. Extended X-ray emission around
the galaxy is detected to a radius of  $\sim8^{\prime}-10^{\prime}$ from the
 core (see also Croston et al. 2008). No correlation is found between
 the X-ray and radio emission on such a large scale.

\subsection{NGC\,383}\label{sec:ngc383}

NGC\,383 is the dominant D galaxy, a part of the Arp~133
system, at the core of the poor group WBL~25, also known as IV Zw
38. It is also a part of the Pisces-Perseus supercluster (an
 optical analysis can be found in Miles et al. 2004).
This galaxy has long been 
known to host the famous, twin-jet FR I source 3C\,31, which belongs to 
the same class of giant radio galaxies as NGC\,315 (Sect.~\ref{sec:ngc315}).
3C\,31 has been studied in detail both on large and small
scales by many authors (e.g., Burch 1977; Blandford \& Icke 1978;
Fomalont et al. 1980; Strom et al. 1983; Andernach et al. 1992; 
Laing \& Bridle 2002; Laing et al. 2008, and references therein).

\begin{figure*}
\centering
\epsscale{1}
\plotone{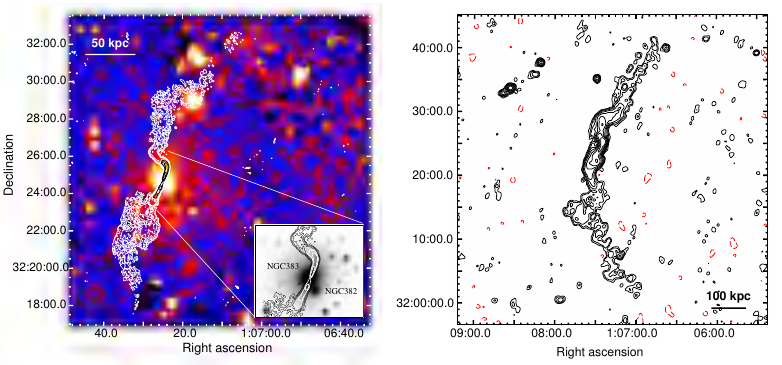}
\caption{NGC\,383. {\em Left:} {\em GMRT} 610 MHz full-resolution contours 
(FWHM=$4.7^{\prime\prime} \times 3.9^{\prime\prime}$, p.a. 
$53^{\circ}$; 1$\sigma$=120 $\mu$Jy beam$^{-1}$) on the smoothed,
0.3-2.0 keV {\em XMM--Newton} image. Contours (black and white) are spaced by a factor 4, 
starting from 0.7 mJy beam$^{-1}$. The inset shows the 
central portion of the 610 MHz radio emission, overlaid on the 
POSS-2 optical image. Contours (black and white) are spaced by a factor 2, starting from 
0.7 mJy beam$^{-1}$. {\em Right:} {\em GMRT} 235 MHz low-resolution image
(FWHM=$38.3^{\prime\prime} \times 34.8^{\prime\prime}$, p.a. 
$-72^{\circ}$;1$\sigma$=1.7 mJy beam$^{-1}$). Black contours are spaced by a 
factor 2 starting from $+3\sigma$. The $-3\sigma$ level is shown as
red dashed contours. 
For this source the scale is 0.346 kpc/$^{\prime\prime}$.  }
\label{fig:ngc383}
\end{figure*}

Fig.~\ref{fig:ngc383} presents our new {\em GMRT} image at 610 MHz 
(left), overlaid on the {\em XMM--Newton} image. The inset shows 
the 610 MHz contours on the optical image of NGC\,383 and its companion 
galaxy NGC\,382, located at $\sim$35$^{\prime \prime}$ ($\sim$ 
12 kpc) south-west of the nucleus. 
The two bright jets undergo multiple bends (wiggles) before 
flowing into elongated, distorted tails. In the right panel of 
Fig.~\ref{fig:ngc383}, we show the low-resolution image at 235 MHz. 
Here the radio tails extend further out and 3C\,31 reaches a size of 
$\sim$ 900 kpc, i.e., the largest scale imaged so far for this source.
Its maximum extent in the images in the literature is $\sim$620 kpc,
as measured at 408 MHz (80$^{\prime \prime}$ resolution; Burch 1977), 
610 MHz (42$^{\prime \prime}$ resolution; Strom et al. 1983), and 1.4
GHz (40$^{\prime \prime}$ resolution; Laing et al. 2008).
Interestingly, the wiggles observed in the jets 
appear to be still occurring even at large distance from the nucleus.

The total flux densities given in Table 4 were measured from images
with  40$^{\prime\prime}$-resolution both at 235 MHz
(Fig.~\ref{fig:ngc383}) and 610 MHz (image not shown here).

The {\em XMM-Newton} image in Fig.~\ref{fig:ngc383} (left) shows 
the X-ray emission from the hot intra-group gas on a scale of 
$\sim$4$^{\prime}$ ($\sim$ 80 kpc) from the central galaxy. The radio 
jets seem to lose collimation and merge into the tails at the boundary 
of the detected X-ray halo. The tails extend well beyond the 
X-ray emission and do not show any evident correlation with the
 X-ray surface brightness (see also Croston et al. 2008). 

\subsection{NGC\,507}\label{sec:ngc507}

NGC~507 (Arp~229) is the dominant elliptical galaxy in the
richest group of galaxies (Mulchaey et al. 2003; O'Sullivan et
al. 2003; Jeltema et al. 2008) in the Pisces system of groups.
It hosts the low-power and extended FR I radio source B2\,0120+33 (Parma et al. 1986). 

\begin{figure*}
\centering
\epsscale{1}
\plotone{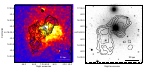}
\caption{NGC\,507. {\em Left:} {\em GMRT} 610 MHz contours 
(FWHM=$7.3^{\prime\prime} \times 5.7^{\prime\prime}$, 
p.a. $60^{\circ}$; 1$\sigma$=50 $\mu$Jy beam$^{-1}$),
overlaid on the smoothed 0.3--2.0 keV {\em Chandra} image. 
Contours are spaced by a factor 2, starting from $+0.25$ mJy 
beam$^{-1}$.  {\em Right:} {\em GMRT} 235 MHz contours
(FWHM=$17.7^{\prime\prime} \times 14.5^{\prime\prime}$, 
p.a. $61^{\circ}$; 1$\sigma$=1 mJy beam$^{-1}$), superposed to 
the POSS-2 image. Radio contours 
are spaced by a factor 2 from $+3\sigma$. For this source the scale 
is 0.336 kpc/$^{\prime\prime}$.}
\label{fig:ngc507}
\end{figure*}

The {\em GMRT} full-resolution image at 610 MHz is presented in Fig.~\ref{fig:ngc507} 
(left) as contours on the {\em Chandra} image. In the right panel, we show 
the overlay of the 235 MHz contours on the optical image. The radio source is 
very distorted and asymmetric. The faint, unresolved central component
in the 610 MHz image is coincident with the nucleus of the optical
galaxy and hosts the radio core, clearly detected at 4.9 GHz (image
from archival {\em VLA} observations -- project AH766 -- not shown
here; see also Murgia et al. 2011). The core is undetected at 235 MHz.
No kpc-scale jets are visible in our images. The eastern 
lobe is more extended, but fainter than the western one. A compression of the 
radio contours in the south-western region of the western lobe suggests a strong 
interaction with the external medium. 

Similar to the radio images, the X-ray emission of NGC\,507 appears strongly
disturbed and asymmetric. The brightest X-ray emission is concentrated in a
$\sim$40 kpc region south-west of the central peak. It surrounds, and
partially overlaps, 
the brightest part of the western radio lobe. The other lobe appears
associated with fainter X-ray emission. A sharp surface brightness edge is 
observed to the east of the core, just beyond the radio emission (Kraft et al.
2004). This discontinuity is also visible in the {\em XMM-Newton} observation 
of the group (Fabbiano, Kim \& Brickhouse 2002) and might be
created by the inflation/expansion of the radio lobe (Kraft et
al. 2004). Tentative evidence for an X-ray cavity at $\sim$7 kpc south-east of
the group center,  i.e., at the base of the western lobe, has been
reported by Dong, Rasmussen \& Mulchaey (2010).

All the X-ray structures described above suggest that the core of
NGC\,507 has been seriously perturbed by the AGN activity in the 
central elliptical.  At the same time, Murgia et al. (2011) suggested
that NGC\,507 is a dying radio galaxy, whose evolution might have been 
significantly affected by the external medium. The fading phase of a radio
galaxy in a particularly dense environment, such as a group or cluster
core, may be longer than that of a field radio source, 
due to the significantly higher pressure of the medium, which is able to confine the 
radio plasma and prevent its dissipation through adiabiatic
expansion (e.g., Slee \& Reynolds 1984; Roland et al. 1985; Murgia et
al. 2011; see also Sect. 6.3).

\subsection{NGC\,741}\label{sec:ngc741}

NGC\,741 is the central and brightest elliptical galaxy of a rich 
fossil group of galaxies at a redshift of $z\!=\! 0.019$ (Mulchaey \& 
Zabludoff 1998; Miles et al. 2004; Forbes et al. 2006).
The velocity dispersion of the group is $\sim 460$ km~s$^{-1}$ (Jetha
et al. 2008), making it one of the richest groups in our sample. There
are $\sim$30 catalogued members of the
group, even though the magnitude difference between the brightest
(NGC\,741) and second brightest (NGC\,742) galaxies is almost three
magnitudes, separated by only $\sim 20^{\prime\prime}$ (17~kpc) in the plane
of the sky. However the velocity 
difference is $\sim400$ km s$^{-1}$, suggesting that NGC\,742 has 
fallen at high velocity through the group core and is passing 
close to NGC\,741. The two galaxies have long been known to be 
associated with the bright, extended radio source 4C 05.10 
(Birkinshaw \& Davies 1985).

\begin{figure*}
\centering
\epsscale{1}
\plotone{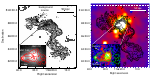}
\caption{NGC\,741. {\em Left panel:} {\em GMRT} 610 MHz full-resolution image of the NGC\,741/742 
system (contours and gray scale;  FWHM= $7.9^{\prime\prime} \times 
4.6^{\prime\prime}$, p.a. $52^{\circ}$; 1$\sigma$=50 $\mu$Jy beam$^{-1}$). 
Contours are spaced by a factor 2, starting from $+0.25$ mJy
beam$^{-1}$. Dashed contours show the $-0.25$ mJy beam$^{-1}$ level.
The inset zooms on the central region (radio contours at 8, 16 and 32 mJy beam$^{-1}$),
overlaid on the POSS-2 optical image. {\em Right panel}: {\em GMRT} 235
MHz full-resolution contours (FWHM=$12.7^{\prime\prime} \times 12.3^{\prime\prime}$, p.a. 
$64^{\circ}$; 1$\sigma$=300 $\mu$Jy beam$^{-1}$), overlaid on the smoothed 
0.3-2.0 keV {\em XMM--Newton} image. Radio contours are spaced 
by a factor 2, starting from 0.9 mJy beam$^{-1}$. The inset shows the 610 
MHz contours on the smoothed, 0.3-2.0 keV {\em Chandra} image. For this source the 
scale is 0.376 kpc/$^{\prime\prime}$.}
\label{fig:ngc741}
\end{figure*}

Our {\em GMRT} full-resolution images at 610 MHz and 235 MHz are presented in 
Fig.~\ref{fig:ngc741}. The radio emission at the group center exhibits 
a very complicated morphology. Two central bright peaks of emission 
appear spatially coincident with the two optical galaxies NGC\,741 and 
NGC\,742 (see inset in the left panel), and are apparently connected by 
a bridge of diffuse emission.
In the south-west, a tail of emission 
extends $\sim150$ kpc from NGC\,741, with a striking twisted structure
and surface brightness decreasing with distance from the 
galaxy. A region of faint emission is located $\sim 25$ kpc east of NGC\,742, with a
total extent of $\sim50$ kpc. The discrete radio source embedded
(in projection) within this emission is likely unrelated to the 
NGC\,741 group, being associated instead with a possible background galaxy. 
No redshift information is available for this object.

It is likely that this peculiar radio structure is the blend of 
two different radio sources. Possible explanations have been 
discussed in the literature (e.g., Venkatesan et al. 1994; 
Birkinshaw \& Davis 1985). In particular, it has been proposed 
that the observed emission is the superposition of a large head-tail, 
associated with NGC\,742, onto a compact source hosted by NGC\,741.
Alternatively, it has been argued that NGC\,742 might be simply
projected by chance on the hot spot of a double radio galaxy 
associated with NGC\,741 (Birkinshaw \& Davis 1985). The 
detection of the eastern diffuse emission in our {\em GMRT} images 
further complicates the interpretation of the source. The origin 
of such a feature is unclear. If the south-west tail originates from 
NGC\,742, then this structure might be emission associated with 
NGC\,741 and bent toward the east by the interaction between the 
two galaxies. An interesting, alternative possibility is that 
the eastern emission is an aged radio lobe inflated by a former outburst 
of the dominant galaxy.

The NGC\,741 system has been observed by both {\em XMM-Newton} and {\em
  Chandra} (Fig.~\ref{fig:ngc741}). The {\em XMM-Newton} observation shows the large-scale
X-ray structure of the group core (Fig.~5, right), while the {\em Chandra} image (inset)
reveals the AGN in both NGC\,741 and its neighbor NGC\,742, as well as narrow
filamentary (or possibly edge-on sheet or shell) structure
linking the two. Interestingly, one of the X-ray filaments appears to run
along the southern border of the diffuse radio bridge between the two
galaxies. An X-ray elliptical cavity, with no radio
emission at frequencies $\ge$ 1.4 GHz, has been detected  by
Jetha et al. (2008) at RA$\approx$ 01h 56m 15s and
DEC$\approx$05$^{\circ}$ 38$^{\prime}$ and 00$^{\prime \prime}$, with
semi-major axes of $\sim 50^{\prime \prime}$ and $\sim 38^{\prime \prime}$.
This ghost cavity has been interpreted as the product of former AGN activity in the
dominant elliptical.  No low-frequency radio emission from this cavity is
detected in our {\em GMRT} images. From the radio and X-ray data, it seems likely
that both AGN outbursts and merger processes have shaped the structures
seen in the group core.

\subsection{HCG\,15}\label{sec:hcg15}

HCG~15 is a compact group of galaxies (also known as WBL66; 
Osmond et al. 2004, Forbes et al. 2006) of six galaxies with a mixed spiral
and elliptical population. It is the only system in our sample that 
appears to be dominated by several bright galaxies, rather than one 
single giant elliptical at the center (Coziol, Brinks \& Bravo-Alfaro 2004). 

\begin{figure*}
\centering
\epsscale{1}
\plotone{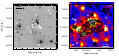}
\caption{HCG\,15. {\em Left:} {\em GMRT} full-resolution image at 610 MHz 
(FWHM=$8.0^{\prime\prime} \times 4.7^{\prime\prime}$, 
p.a. $63^{\circ}$; 1$\sigma$=35 $\mu$Jy beam$^{-1}$), overlaid on the 
POSS-2 optical image. 
A and B indicate the point sources 
associated with HCG\,15d and HCG\,15a,  respectively. 
{\em Right:} {\em GMRT} 610 MHz low-resolution contours (black; 
FWHM=$23.8^{\prime\prime} \times 20.0^{\prime\prime}$, p.a. $71^{\circ}$
, 1$\sigma$=40 $\mu$Jy beam$^{-1}$), overlaid on the smoothed 0.3-2.0 keV 
{\em XMM--Newton} image. In both panels the radio contours are 
spaced by a factor 2, starting from +3$\sigma$. The $-3\sigma$ level
is shown as dashed contours. For this source the scale is 0.460 kpc/$^{\prime\prime}$.}
\label{fig:hcg15}
\end{figure*}

The  {\em GMRT} 610 MHz full-resolution image is presented in 
Fig.~\ref{fig:hcg15} (left) as contours overlaid on the optical image. 
Two point sources (labelled A and B in the figure) are associated with
the group galaxies HCG\,15d and 
HCG\,15a (Coziol, Brinks \& Bravo-Alfaro 2004), respectively.
In addition to these sources,  low-brightness, extended 
emission is found in the region south of source A, as emphasized by the
low-resolution 610 MHz image shown in the right panel (contours). 
The diffuse source appears clumpy and elongated in the northwest-southeast, with a total 
extent of $\sim$150 kpc. The {\em GMRT} observation at 235 MHz was affected by 
severe RFI.  After considerable flagging and many cycles of phase self-calibration, 
the quality of the final images still appears compromised by residual RFI and 
phase errors. For this reason, we do not show the 235 MHz image.
A deeper re-observation at 235 MHz (scheduled during {\em GMRT} 
Cycle 18) will provide a better image of the group emission at this frequency.  

The {\em XMM-Newton} X-ray observation of HCG\,15 reveals disturbed, 
diffuse emission between the group galaxies, particularly on the eastern 
side (Fig.~\ref{fig:hcg15}, right; see also Finoguenov et al. 2007),
suggesting that the system is not in a fully relaxed state.
The brightest emission is centered on HCG\,15d (A).
The second brightest region is a clump of emission, with an extent of
$\sim$40 kpc, between HCG\,15a (B) and the rest of the group. The diffuse 
radio source appears associated with the intra-group medium rather than with
the individual galaxies, and extends out to the X-ray clump on the
eastern side of the group. This poses the question of its origin. The
extended radio emission may be an old,
detached radio lobe which buoyantly moved away from the host galaxy, possibly 
subject to the intra-group gas motion. An alternative scenario is suggested by the 
analogy with the case of Stephan's Quintet, where a bridge of X-ray and 
radio emission crosses the center of the group, interpreted as 
material shock-heated by the collision of an interloper galaxy 
(O'Sullivan et al. 2009, and references therein). This raises the 
interesting possibility that the diffuse radio emission in HCG\,15 might 
arise from a broad shock front caused by the passage of a galaxy through the 
group core. In this scenario, we might expect discordant galaxy velocities and 
a high galaxy velocity dispersion. 
The radial velocities of the six principal galaxies are spread over $\sim1000$ 
km s$^{-1}$, and the dispersion is $\sim$400 km s$^{-1}$ (Hickson,
Kindl \& Auman 1989; Mulchaey et al. 2003).

\subsection{NGC\,1407}\label{sec:ngc1407}

NGC\,1407 is the dominant elliptical galaxy in a dynamically evolved
rich group (Eridanus A) dominated by dwarf galaxies (e.g., Miles et al.
2004; Trentham et al. 2006), and it is part of a larger system of
groups called the Eridanus super-group (Brough et al. 2006).

\begin{figure*}
\centering
\epsscale{1}
\plotone{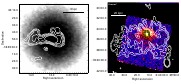}
\caption{NGC\,1407. {\em Left:} {\em GMRT} 610 MHz full-resolution contours 
(FWHM=$5.6^{\prime\prime}\times4.3^{\prime\prime}$, p.a. $41^{\circ}$;
1$\sigma=$ 0.10 mJy beam$^{-1}$) on the POSS-2 red optical image. 
Contours are spaced by a factor 2 starting from $+3\sigma$. 
{\em Right:} {\em GMRT} 235 MHz low-resolution contours (FWHM=$48.5^{\prime\prime} \times 
31.9^{\prime\prime}$, p.a. $8^{\circ}$; 1$\sigma=$ 1 mJy beam$^{-1}$), overlaid
on the smoothed,  0.3-2.0 keV {\em Chandra} image. Contours are spaced by a 
factor 2, starting from +3$\sigma$. The black contour region approximately
corresponds to the region occupied by the 610 MHz contours in the left
panel.
 For this source the scale 
is 0.122 kpc/$^{\prime\prime}$. }
\label{fig:ngc1407}
\end{figure*}

In Fig.~\ref{fig:ngc1407} (left), we present the 610 MHz 
full-resolution contours on the optical image. 
The radio source has an asymmetric twin-jet structure, with a 
total size of only $\sim$6 kpc. The eastern jet broadens at
$\sim$2 kpc from the central radio peak, while, at a similar 
projected distance, the western jet undergoes a sharp bend toward the north.
The 235 MHz image (white and black contours), at a resolution of 
$49^{\prime \prime} \times 32^{\prime \prime}$, is shown in the right
panel of Fig.~\ref{fig:ngc1407}, 
overlaid on the {\em Chandra} image. The black contour region approximately
corresponds to the region covered by the 610 MHz contours.
The emission at this frequency is one order of magnitude more 
extended than the emission detected at 610 MHz. A very 
diffuse and low-brightness structure of $\sim$80 kpc extent completely 
encloses the higher frequency emission. No indication of this diffuse 
component is found in low resolution images at 610 MHz, though the 
brightest peaks are marginally visible (at the 1$\sigma$ level) in the 
1.4 GHz NVSS image. The flux measured at 235 MHz for this diffuse component is 
1170 mJy, while the NVSS gives $S_{\rm 1.4 \, GHz}\sim40$ mJy,
implying a very steep spectral index ($\alpha \sim 1.9$). 
Our images suggest that two distinct radio outbursts co-exist in this
system: the current AGN activity, 
which can be identified with the small-scale double radio source, 
and the diffuse component, generated by an earlier outburst.
The hypothesis of a restarted radio galaxy is consistent with
the ultra-steep spectrum of the source (Table 4), which is
dominated by the low-frequency diffuse component. 

The {\em Chandra} image in Fig.~\ref{fig:ngc1407} (right) shows that
the X-ray emission of the group is mostly extended along the east-west 
axis, and more extended to the south than the north. The 235
MHz diffuse component extends on a larger scale ($\sim$80) than the X-ray
halo ($\sim$ 30kpc). However, the radio structure shows a similar north-south
asymmetry, suggesting a common cause, probably motion of the group 
northward. Using {\em Chandra} data, Dong, Rasmussen \& Mulchaey (2010) 
located a possible X-ray cavity at less than 0.5 kpc from the X-ray
center of the group. The resolution of our 610 MHz image does not
allow a direct comparison between the radio emission and the candidate
cavity, which lies in the region covered by the highest radio contour
in Fig.~\ref{fig:ngc1407} (left).

\subsection{NGC\,1587}\label{sec:ngc1587}

NGC\,1587 is the brighter of the pair of elliptical galaxies NGC\,1587/88,
which is part of a poor group with an unusually low velocity dispersion
$\sim$100 km~s$^{-1}$ (Mulchaey et al. 2003, Osmond \& Ponman 2004).

\begin{figure*}
\centering
\epsscale{1}
\plotone{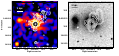}
\caption{NGC\,1587. {\em Left}: {\em GMRT} 610 MHz full-resolution 
contours (FWHM=$5.7^{\prime\prime} \times 4.7^{\prime\prime}$, p.a. 
$64^{\circ}$; 1$\sigma$=50 $\mu$Jy beam$^{-1}$), overlaid 
on the smoothed 0.3-2.0 keV {\em Chandra} image. {\em Right}: {\em GMRT} 
610 MHz low-resolution 
contours (FWHM=$12.0^{\prime\prime} \times 10.0^{\prime\prime}$, p.a. $0^{\circ}$;
1$\sigma$=120 $\mu$Jy beam$^{-1}$), overlaid on the POSS-2 red optical image.  
In both panels contours are spaced by a factor 2, starting from the $+3\sigma$ 
level. For this source the scale is 0.252 kpc/$^{\prime\prime}$.}
\label{fig:ngc1587}
\end{figure*}

The 610 MHz radio image of NGC\,1587 is shown 
in Fig.~\ref{fig:ngc1587} at full resolution (left) and slightly lower 
resolution (right), overlaid on the {\em Chandra} and POSS-2 images, 
respectively. The source has a compact, 
bright component, coincident with the galaxy optical peak. 
This component appears surrounded by amorphous, low-brightness 
emission that encompasses a $\sim$5 kpc radius region around the 
center. A fainter structure extends for $\sim$10 kpc 
north-west of the galaxy. No radio emission is detected from the
companion galaxy NGC\,1588. The radio 
morphology of NGC\,1587 is reminiscent of core-halo radio sources,
found in a number of clusters with 
cool cores, such as 3C\,317 (Venturi, Dallacasa \& Stefanachi 2004),
2A\,0335+096 (Sarazin, Baum \& O'Dea 1995), PKS\,0745-191 
(Baum \& O'Dea 1991), RXJ\,1720.1+2638 and MS\,1455.0+2232 (Mazzotta
\& Giacintucci 2008). 
However,
core-halos usually extend over a scale which is comparable to the core
of the cluster (i.e., from several tens to few hundreds of kpc in
radius), while the halo component in NGC\,1587  appears fully
contained within the optical envelope of the galaxy.
Our {\em GMRT} observation at 235 MHz detected only the central 
point-source. For this reason, we do not show the 235 MHz
image, and we only provide the total flux density in Table 4. 

The {\em Chandra} image is shown in Fig.~\ref{fig:ngc1587} (left). The group 
has very faint extended X-ray emission  
and low temperature (kT=0.4 keV; Helsdon et al. 2005). The radio emission 
appears associated with the X-ray brightest regions, with the exception of the faint 
radio feature extended toward the north-west. Helsdon et al. (2005) 
found an indication of a gas temperature peak in the central
1$^{\prime}$ radius 
region of the group. The radio source seems to be totally contained 
within this radius, suggesting that the gas might have been
heated by its activity.

\subsection{MKW\,2}\label{sec:mkw02}

MKW\,2 is a rich group (or poor cluster, also known as WBL\,276), centered on the
cD galaxy CGCG\,009-062. It is part of a complex system, with a poor group
about 2 Mpc from its center, projected on the sky to its north-east, and
another equally rich system (MKW\,2s) towards the west, in the foreground.

The {\em GMRT} 235 and 610 MHz observations re-analyzed here are 
from Giacintucci et al. (2007), who observed the central galaxy as 
part of a study of the radio emission associated with cD galaxies in rich and 
poor clusters. The cD in MKW\,2 hosts a very large radio 
galaxy with a morphology similar to that of FR II sources, with jets, 
asymmetric in size and brightness, lobes and possibly hot spots (Burns et al. 1987;
 Giacintucci et al. 2007), though its radio power is typical of FR I radio galaxies (Table 1). 

\begin{figure*}
\centering
\epsscale{1}
\plotone{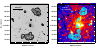}
\caption{MKW\,02. {\em Left}: {\em GMRT} 610 MHz contours (FWHM=$25.8^{\prime\prime} \times 18.4^{\prime\prime}$, 
p.a. $28^{\circ}$; 1$\sigma$=200 $\mu$Jy beam$^{-1}$) on the POSS-2
red optical image. {\em Right}: {\em GMRT} 235 MHz contours
(FWHM is $17.8^{\prime\prime} \times 13.9^{\prime\prime}$, p.a. 
$42^{\circ}$; 1$\sigma$=500 $\mu$Jy beam$^{-1}$), overlaid on the
smoothed 0.3--2.0 keV {\em XMM--Newton} image.  In both panels contours are 
spaced by a factor 2, starting from the
$+3\sigma$ level. For this source the scale is 0.731 kpc/$^{\prime\prime}$.}
\label{fig:mkw2}
\end{figure*}

In Fig.~\ref{fig:mkw2} we show new images at 610 MHz (left) and 
235 MHz (right) produced with a lower resolution with respect to
the images in Giacintucci et al. (2007; see their Fig.~11). The radio contours are overlaid 
on the optical and {\em XMM-Newton} images. As in Giacintucci et al. (2007), only the 
central component, coincident with the galaxy, and the radio lobes
are detected at 610 MHz. The southern jet is completely resolved out, 
while a possible knot of the northern jet is visible at $\sim$ 100 kpc from the 
center. The 235 MHz image shown here reveals more structure 
than the higher resolution image in Giacintucci et al. (2007), since the southern jet is now
almost entirely detected, as well as part of the other jet beyond
the knot.

The total flux  measured from the new {\em GMRT} images in Fig.~\ref{fig:mkw2} 
is 1.95 Jy at 235 MHz and 674 mJy at 610 MHz (Table 4). These values are 
higher than the fluxes given in Giacintucci et al. (2007), 1.7 Jy and 245 mJy, respectively. 
While the difference at 235 MHz is $\sim$15\% and might be due to the 
different angular resolution of the images, the disagreement between the
610 MHz fluxes is much larger. By comparison with the NVSS flux density 
(Table 1) and literature data, Giacintucci et al. (2007) inferred that their 610 MHz flux was 
most likely underestimated.

Fig.~\ref{fig:mkw2} (right) presents the X-ray image of MKW\,2 from new
{\em XMM-Newton} observations obtained during cycle 5.  Previous {\em ROSAT PSPC}
observations (e.g., Mulchaey et al 2003) suggested an interaction between the
lobes and the intra-cluster medium.  The new {\em XMM-Newton} data do not
show a clear correlation, probably owing to the very low density of
  the gas in the group and the distance of the radio lobes from the
  cluster center.  Note that the northern lobe falls partially on the
EPIC-MOS chip gaps. The knots of the jets seem to be roughly
coincident with some bright features in the X-ray image. 

\subsection{NGC\,3411}\label{sec:ngc3411}

The large elliptical galaxy NGC\,3411 lies at the center of a relaxed
group with several neighboring disk galaxies. The dominant galaxy is sometimes
misidentified as NGC\,3402.  Previous high-resolution ($<4^{\prime
  \prime}$) {\em VLA} observations at 1.4 GHz, 4.9 GHz and 8.4 GHz detected a resolved
nuclear source and marginally extended emission with no indication of any jets
(O'Sullivan et al. 2007). The comparison with the 1.4 GHz
NVSS image suggested the existence of larger scale, low-surface brightness
emission which was not detected in those observations.

\begin{figure*}
\centering
\epsscale{1}
\plotone{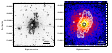}
\caption{NGC\,3411. {\em Left:} {\em GMRT} 610 MHz full-resolution contours 
(FWHM=$6.9^{\prime\prime} \times 5.3^{\prime\prime}$, 
p.a. $-19^{\circ}$; 1$\sigma$=90 $\mu$Jy beam$^{-1}$), overlaid on the 
POSS-2 optical image. 
Contours are spaced by a factor 2, starting from $+3\sigma$. 
The $-3\sigma$ level is shown as dashed contours.
 {\em Right:} {\em GMRT} 235 MHz contours (FWHM=$18.0^{\prime\prime} 
\times 15.0^{\prime\prime}$, p.a. $0^{\circ}$; 1$\sigma$=400 $\mu$Jy 
beam$^{-1}$), overlaid on the smoothed 0.3-2.0 keV {\em Chandra} image.
Contours are spaced by a factor 2, starting from the 3$\sigma$ level.
For this source the scale is 
0.312 kpc/$^{\prime\prime}$.}
\label{fig:ngc3411}
\end{figure*}

Our {\em GMRT} images are presented in Fig.\ref{fig:ngc3411}. On the left we 
show the 610 MHz emission at the full resolution, overlaid on the POSS-2.
The right panel presents the 235 MHz contours at the resolution of 
$18^{\prime \prime} \times 15^{\prime \prime}$, overlaid
on the {\em Chandra} image. The low-frequency morphology of NGC\,3411 is 
characterized by a diffuse component, mostly extended in the
north-south direction, 
which covers a much larger scale ($\sim$80 kpc) than the high frequency 
images in O'Sullivan et al. (2007). Such emission appears as an irregular 
halo surrounding the central compact component. No jets connecting 
the point source and the extended emission are visible in our images. 
This morphology resembles the core-halo radio sources typically
found at the centers of relaxed cool-core clusters (e.g., 
Mazzotta \& Giacintucci 2008 and references therein; see also
Sect.~\ref{sec:ngc1587}), although the total extent of the source in
NGC\,3411 is smaller than the typical size of core-halos (from several 
tens to few hundreds of kpc in radius).
The source has a steep spectral index ($\alpha = 1.44$) -- one of
the steepest among the sample (Table 4) -- as usually found for
core-halo sources.

The X-ray emission of the NGC\,3411 group appears completely relaxed 
(Fig.~\ref{fig:ngc3411}, right). However, O'Sullivan et al. (2007) found a highly unusual
temperature structure in the profiles and temperature maps from
both {\em Chandra} and {\em XMM-Newton} data; the system 
has a hot core surrounded by a shell of cooler gas extending between
$\sim$20 kpc and 40 kpc from the center. O'Sullivan et al. (2007) argued that NGC\,3411 
could be an example of a re-heated cool core, with a recent AGN
outburst heating the central cool gas to the observed temperature.

\subsection{NGC\,4636}\label{sec:ngc4636}

NGC\,4636 is the dominant elliptical of a nearby poor group on the
periphery of the Virgo cluster. The kinematics of globular clusters
reveal an extended dark matter halo for the dominant elliptical (Chakrabarty
\& Raychaudhury 2008), in agreement with mass estimates from 
{\em Chandra} and {\em XMM-Newton} observations (Johnson et al. 2009).

\begin{figure*}
\centering
\epsscale{1}
\plotone{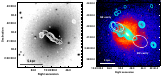}
\caption{NGC\,4636. {\em Left:} {\em GMRT} 610 MHz full-resolution contours
(FWHM=$5.8^{\prime\prime} \times 4.3^{\prime\prime}$, p.a. $48^{\circ}$;
1$\sigma$=50$\mu$Jy beam$^{-1}$), overlaid on the POSS-2 optical image. 
Contours are spaced by a factor 2, starting from $+3\sigma$. The
$-3\sigma$ level is shown as dashed contours.
{\em Right:} {\em GMRT} 235 MHz full-resolution contours (FWHM=$15.7^{\prime\prime} \times 
12.9^{\prime\prime}$, p.a. $31^{\circ}$; 1$\sigma$=170 $\mu$Jy beam$^{-1}$),
overlaid on the smoothed 0.3--2.0 keV {\em Chandra} image. Contours are 
spaced by a factor 2, starting from $+3\sigma$. For this source the scale is 
0.064 kpc/$^{\prime\prime}$.}
\label{fig:ngc4636}
\end{figure*}

Our {\em GMRT} image at 610 MHz has been presented in 
Baldi et al. (2009b), who investigated the X-ray morphology and temperature structure of the group core 
using a long {\em Chandra} exposure. Here we report the 610 MHz image, 
overlaid on the POSS-2, in the left panel of Fig.~\ref{fig:ngc4636}, 
and present the new 235 MHz contours superimposed on the {\em Chandra}
image in the right panel.

The radio source structure is visible in better detail in the high-resolution 
image at 610 MHz. The source has a small double morphology, with bright jets 
flowing into fainter radio lobes that extend to a radius of 
$\sim$2.5 kpc from the center. The overall radio structure is symmetric in 
brightness and size, and appears twisted in a peculiar, reversed S-shape. 
The double radio source to the east is a background radio galaxy. At lower 
frequency (Fig.~\ref{fig:ngc4636}, right) the source becomes asymmetric;  
both lobes are more extended than at 610 MHz, but the emission on the north-east 
side reaches a larger distance ($\sim$7 kpc) from the center than the south-west
($\sim$ 3.5 kpc).

The {\em Chandra} images (Fig.~\ref{fig:ngc4636}, right; Jones et al.
2002; Ohto et al. 2003; O'Sullivan, Vrtilek \& Kempner  2005; Baldi et
al. 2009a) reveal a very unusual morphology in the core of
NGC\,4636. The most prominent feature is a set 
of quasi-symmetric, spiral arm-like structures (only barely visible in
Fig.~\ref{fig:ngc4636}; see Baldi et al. 2009a for more detailed
images of these features), which are coincident with 
the border of two X-ray cavities (NE cavity and SE cavity in
Fig.~\ref{fig:ngc4636}). This complex X-ray morphology is believed to be produced by 
AGN-driven shocks (Jones et al. 2002). 
The presence of 610 MHz radio emission extending towards the bubbles 
might indeed confirm such a scenario (Baldi et al. 2009a). Further confirmation 
comes from the 235 MHz image in Fig.~\ref{fig:ngc4636}, which shows the 
north-east cavity to be totally filled with low-frequency radio emitting plasma. On the 
opposite side, the south-west cavity remains undetected in the radio,
except for a small overlap 
with the 235 MHz emission that points toward the center of the cavity. 
The combination of the radio and X-ray data indicates that NGC\,4636
is a rather complex and intriguing multi-bubble system, with all the
cavities, some radio-loud and other radio-quiet, located at roughly
the same distance from the center. 

\subsection{HCG\,62}\label{sec:hcg62}

HCG\,62 is one of the nearest Hickson compact galaxy groups (Hickson, Kindl \& Auman 1989)
and the brightest group of galaxies in the X-ray band (Ponman and Bertram 1993). 
It contains $\sim$ 60 galaxies within a radius of approximately 900 kpc (Mulchaey et 
al. 2003). The central galaxy hosts an extended, low-power radio
source (Vrtilek et al. 
2002). 

\begin{figure*}
\centering
\epsscale{1}
\plotone{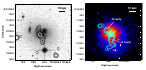}
\caption{HCG\,62. {\em Left:} {\em GMRT} 610 MHz contours
(FWHM=$14.0^{\prime\prime} \times 14.0^{\prime\prime}$, p.a. 
$0^{\circ}$; 1$\sigma$=50$\mu$Jy beam$^{-1}$), overlaid 
on the POSS-2 optical image. Contours are spaced by a factor 
2, starting from $+3\sigma$. {\em Right:} {\em GMRT} 235 MHz contours
(FWHM=$14.1^{\prime\prime} \times 12.3^{\prime\prime}$, p.a. 
$47^{\circ}$; 1$\sigma$=170 $\mu$Jy beam$^{-1}$) on
on the smoothed, 0.3--2.0 keV {\em Chandra} image. Contours 
are spaced by a factor 2, starting from $+3\sigma$. 
For this source the scale is 
0.280 kpc/$^{\prime\prime}$.}
\label{fig:hcg62}
\end{figure*}

Our {\em GMRT} images at 235 MHz and 610 MHz have been presented in Gitti et al.
(2010). Here we report the images in Fig.~\ref{fig:hcg62}, overlaid on the POSS-2 and
smoothed {\em Chandra} images. 
The source has a double morphology at both frequencies. 
At 610 MHz the radio lobes extend out to a radius of $\sim$14 kpc from the center. 
At lower frequency, the emission reaches a larger distance ($\sim$22 kpc) and 
appears bent toward the east. 

HCG\,62 represents one of the clearest cases of AGN interaction in our sample. 
The {\em Chandra} image in Fig.~\ref{fig:hcg62} shows the two prominent
X-ray cavities detected to the north and south of the group core by
Vrtilek et al. (2002). This was the first report of cavities in a group of galaxies. 
Previous radio observations at 1.4 GHz revealed only weak extended
emission, partially overlapping the southern cavity, and failed to
detect any radio emission from the northern cavity (Vrtilek et
al. 2002). Gitti et al. (2010) used  {\em Chandra} and
{\em XMM-Newton} data in combination with the {\em GMRT} images shown in this paper
to study the X-ray properties of the hot gas in HCG\,62 and its interaction
with the central radio galaxy. Both cavities were found to be associated with the
low-frequency radio plasma detected by our {\em GMRT}
observations. The radio lobes totally fill the cavities at both
frequencies, and even extend beyond them at 235 MHz (Fig.~\ref{fig:hcg62},
right). The presence of such faint radio emission outside the cavities, with
no counterpart at 610 MHz,  may suggest two distinct episodes of AGN
outbursts in this group. Deeper multi-frequency {\em GMRT}
observations, carried out during  Cycle 17, will allow us to test
this possibity.

\subsection{NGC\,5044}\label{sec:ngc5044}

NGC\,5044 is the central member of a rich group, which is the brightest
in X-ray luminosity in the GEMS compilation of nearby groups (Osmond \&
Ponman 2004, Miles et al. 2004, 2006). The radio data
available in the literature prior to this work, and in the {\em VLA}
archives, are limited to a few observations that detect only a flat-spectrum 
radio core (e.g., Sparks et al. 1984). 

\begin{figure*}
\centering
\epsscale{1}
\plotone{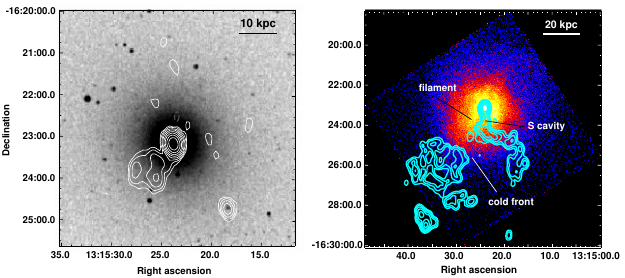}
\caption{NGC\,5044. {\em Left:} {\em GMRT} 610 MHz low-resolution
contours (HPBW=$18.3^{\prime\prime} \times 12.7^{\prime\prime}$, 
p.a. $22^{\circ}$; 1$\sigma$=75 $\mu$Jy beam$^{-1}$), overlaid on 
the POSS-2 red optical image. Contours are spaced by a factor 2, 
starting from $+3\sigma$. The $-3\sigma$ level is shown as dashed contours.
{\em Right:} {\em GMRT} 235 MHz low-resolution contours (HPBW=$22.0^{\prime\prime} \times 
16.0^{\prime\prime}$, p.a. $0^{\circ}$; $1\sigma=$250 $\mu$Jy beam$^{-1}$),
overlaid on the smoothed, 0.3--2.0 keV {\em Chandra} image. Contours are 
spaced by a factor 2, starting from +3$\sigma$. 
For this source the scale is 0.185 kpc/$^{\prime\prime}$.}
\label{fig:ngc5044}
\end{figure*}

Our {\em GMRT} images at 235 MHz and 610 MHz have been presented in David et al.
(2009). Here we report the images in Fig.~\ref{fig:ngc5044} overlaid on the optical (left) and {\em
  Chandra} (right) images. Beyond the central core, extended emission is clearly revealed by
our {\em GMRT} observations; a faint radio lobe, with a size of $\sim$12 kpc, is
detected at 610 MHz. The radio emission at 235 MHz
is much more extended than the emission at 610 MHz, with little
overlap between the two frequencies. The 235 MHz image shows a bright 
radio jet, misaligned with respect to the 610 MHz lobe. The jet is not straight: a
bend occurs toward the west at $\sim$9 kpc from the core; at a larger
distance, the jet deviates again, curving by almost 90 degrees toward the
south. The 235 MHz image also reveals a region of low-surface brightness emission,
with a rather clumpy appearance, located at $\sim2.5^{\prime}$ ($\sim$28
kpc) south-east of the core and apparently detached from the radio
jet. The lack of any detected emission at 610 MHz in the same region
as the 235 MHz emission indicates that the emission must be very steep
with a spectral index of $\alpha >1.6$. 

The deep {\em Chandra} observation of NGC\,5044 analyzed by David et
al. (2009) revealed interesting X-ray substructures, clearly visible in both the
raw and unsharp mask images (their figures 1 and 3 ), and most 
likely generated by the interaction of the group gas with the central 
radio galaxy. The axis of the 610 MHz lobe was found to be coincident
with a filament of cool gas extending approximately 20 kpc to the
south-east (see also Fig.~\ref{fig:ngc5044}, right). This was 
interpreted as gas being lifted from the group 
center by the buoyantly rising lobe, which is being inflated by an ongoing AGN 
outburst in NGC\,5044. A number of small X-ray cavities were 
found around the group center (see also Gastaldello et al. 2009).
No radio emission is detected from these cavities at the sensitivity 
level of our {\em GMRT} observations (Table 1), with the exception of the 
largest cavity, south of the center, which is totally filled by the 
the inner part of the 235 MHz jet (right panel of Fig.~\ref{fig:ngc5044}; 
see also Figs. 5 and 6 in David et al. 2009). Interestingly, the detached 
radio lobe, detected at 235 MHz, appears to be located just beyond 
a cold front at $\sim$30 kpc toward the south-east from NGC\,5044
(David et al. 2009, 2011).
Most likely, the 235 MHz emission was produced by past activity of
the central AGN. It is unclear whether the jet and detached lobe 
come from the same outburst or possibly two separated outbursts.
New, deeper {\em GMRT} observations at 235, 327 and 610 MHz (carried out
during  Cycle 17) will allow us to study with unprecedented detail the
complex and unique low-frequency radio properties of NGC\,5044.

\subsection{NGC\,5813}\label{sec:ngc5813}

The giant elliptical NGC\,5813 dominates a group of early-type 
galaxies. The system is a subgroup of a larger complex that
includes the more massive group NGC\,5846 (Sec.~\ref{sec:ngc5846}),
which lies at $\sim$1.4$^{\circ}$ from NGC\,5813 in the plane of the sky
(Mahdavi, Trentham \& Tully 2005; Eigenthaler \& Zeilinger 2010).
With a total radio power of $1.4\times 10^{21}$ W Hz$^{-1}$ at 
1.4 GHz, the radio source associated with NGC\,5813 is the least powerful 
in our sample (Table 1).  

\begin{figure*}
\centering
\epsscale{1}
\plotone{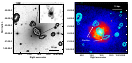}
\caption{NGC\,5813. {\em GMRT} 235 MHz contours (FWHM=$16.0^{\prime\prime} \times 
13.0^{\prime\prime}$, p.a. $0^{\circ}$; 1$\sigma$=300 $\mu$Jy beam$^{-1}$),
overlaid on the POSS-2 optical image (left) and the smoothed, 0.3--2.0 keV
{\em Chandra} image (right). Contours are spaced by a factor 2, starting from 0.9 mJy 
beam$^{-1}$. The inset in the left panel shows the 5$^{\prime \prime}$-resolution
1.4 GHz image from the FIRST (gray scale and contours). For this source the 
scale is 0.135 kpc/$^{\prime\prime}$.}
\label{fig:ngc5813}
\end{figure*}

We observed NGC\,5813 only at 235 MHz (Fig.~\ref{fig:ngc5813}). The galaxy hosts a central,  
triple component $\sim$9 kpc in extent. The 1.4 GHz image from the 
Faint Images of the Radio Sky at Twenty-Centimeters (FIRST) resolves
this component into a point source and two faint extended
structures (inset).  At larger distances from the center, the 235 MHz
image reveals a  pair of lobes located along the same axis of the central component. The northern lobe 
appears detached, whereas the southern lobe is closer and apparently 
connected to the central region. The overall morphology of NGC\,5813
may recall the double-double radio galaxies (e.g., Schoenmakers
et al. 2000), 
consisting of a pair of double lobes having a common nuclear source,
although the linear size of double-double sources is usually much
larger than NGC\,5813. Both outer lobes in NGC\,5813 are undetected in 
the 1.4 GHz NVSS image, which shows only a point source coincident
with the triple component detected at 235 MHz. This implies a very
steep spectral index ($\alpha > 2$) for both lobes.

In the X-ray band, the NGC\,5813 group possesses a system of multiple,
aligned cavities at 1 kpc, 8 kpc and 20 kpc from the central source
(Randall et al. 2011).  The smoothed {\em Chandra} image is shown in Fig.~\ref{fig:ngc5813}
(right). Two small,  symmetrical X-ray cavities with bright rims
are located at only $\sim$1 from the centre (Randall et al. 2011). These inner cavities, 
not visible in  Fig.~\ref{fig:ngc5813} due to the smoothing, are
associated with the extended radio emission visible 
in the FIRST image.  Fig.~\ref{fig:ngc5813} shows a second pair of larger cavities  at $\sim$8 kpc 
from the centre, along the same axis as the inner bubbles. These appear as ghost cavities at 
1.4 GHz, but are clearly filled by the radio lobes detected at 235 MHz. The {\em Chandra} data
also reveal the existence of an additional third set of cavities at a
radius of $\sim$ 20 kpc, nicely aligned 
with the other two pairs (Randall et al. 2011). The combination of the radio and 
X-ray properties of this system suggests multiple episodes of activity: the inner cavities 
might be associated with the most recent, central radio activity; the steep-spectrum 
235 MHz radio bubbles may have been inflated by a former AGN outburst. Indeed,
double-double radio galaxies are considered one of the most striking examples of 
recurrent activity, when a new pair of inner lobes are produced close to the nucleus 
before the previously generated, more distant lobes have completely faded (e.g., Subrahmanyan,
Saripalli \& Hunstead 1996; Lara et al. 1999; see also Saikia \&
Jamrozy 2009 for a recent review on recurrent activity in
AGNs). Finally, the outermost cavities might be associated with even
older radio activity, undetected at the
sensitivity limit of the existing 235 MHz image. 
Deeper {\em GMRT} observations at 235, 327 and 610 MHz have been
obtained during Cycle 18 to investigate in detail the recurrent activity in this system.

\subsection{NGC\,5846}\label{sec:ngc5846}

NGC\,5846 is the giant, dominant elliptical of a massive and rather 
isolated galaxy group, which has more than 250 possible member galaxies 
with spectroscopic confirmation for $\sim$80 of them (Mahdavi et
al. 2005, Forbes et al. 2006). 
The galaxy has very low-power radio emission at 1.4 GHz, and, in fact, it 
is the second weakest source in Table 1.

\begin{figure*}
\centering
\epsscale{1}
\plotone{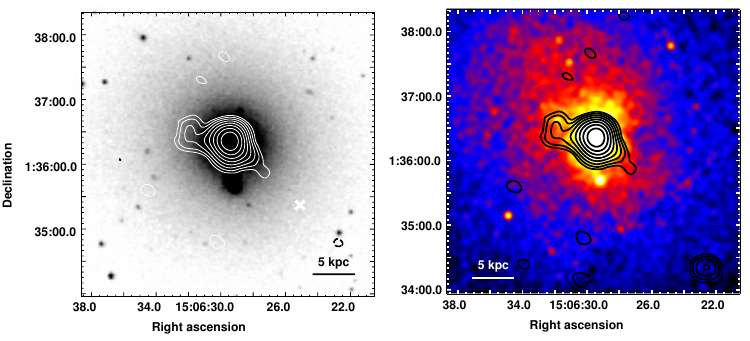}
\caption{NGC\,5846. {\em GMRT} 610 MHz contours (FWHM=$15.0^{\prime\prime} 
\times 15.0^{\prime\prime}$, p.a. $0^{\circ}$; 1$\sigma$=40 $\mu$Jy beam$^{-1}$),
overlaid on the optical POSS-2 image (left) and smoothed, 0.3--2.0
keV {\em Chandra} image (right). Contours are spaced by a factor 2, starting from  
120 $\mu$Jy beam$^{-1}$. 
For this source the scale is 0.118 kpc/$^{\prime\prime}$.}
\label{fig:ngc5846}
\end{figure*}

We observed the radio galaxy with the {\em GMRT} only at 610 MHz. 
During the observation, several antennas suffered
problems and were flagged during the data reduction.
 This led to a rather asymmetric beam (15$^{\prime \prime} 
\times 6^{\prime \prime}$; Table 2) in the images produced from the 
full array. In Fig.~\ref{fig:ngc5846}, we present the 610 MHz image 
convolved with a circular beam with FWHM=15$^{\prime \prime}$. At this 
resolution, we detect an unresolved, bright component, and a small extension 
of $\sim$3 kpc toward the east. Hints of much fainter extended
emission are visible on the opposite side, though it is only one
contour level at $3\sigma$ significance. 

In the X-ray band, the NGC\,5846 group has been studied in detail by 
both {\em Chandra} (e.g., Trinchieri \& Goudfrooij 2002) and {\em XMM-Newton} 
(e.g., Finoguenov et al. 2006) observations. On the large scale, the group 
has a very asymmetric X-ray halo, whose central part is shown in Fig.~\ref{fig:ngc5846} 
(right). {\em Chandra} also revealed a complex X-ray morphology in the 
innermost region (radius $<2$ kpc) of the group core. The main feature is 
a prominent, arc-like structure (Trinchieri \& Goudfrooij 2002). 
Two small X-ray cavities have been identified at $< 1$ kpc from the
group center by \cite{allen06} and confirmed by Dong, Rasmussen
\& Mulchaey (2010). The low resolution of our 610 MHz image does not
allow a direct comparison between the radio emission and these X-ray
cavities, which are all located in the region covered by the radio point source. 
Recently, Machacek et al. (2011) found an X-ray cavity $\sim 7$ kpc 
west of the central source. No radio emission is detected at 610 MHz 
in this cavity.

\subsection{AWM\,4}\label{sec:awm4}

AWM\,4 is a poor cluster of galaxies centered on the giant elliptical 
NGC\,6051 which hosts the radio galaxy 4C+24.36. Our {\em GMRT} images
at 235, 327 and 610 MHz have been presented and analyzed in
Giacintucci et al. (2008). Here we report the 235 MHz and 610 MHz flux 
densities and corresponding spectral index in Table 4. 
Giacintucci et al. (2008) classified 4C+24.36 as a FR I/FR II transition 
wide-angle-tail (WAT), where both ram pressure, caused by the motion
southward of the host galaxy, and buoyancy forces seem to have shaped 
the radio structure. The source possesses twin jets with prominent,
symmetric wiggles, and large lobes extending $\sim$80 kpc from the 
radio core. Based on the jet to counterjet brightness ratio, the
central $\sim$10 kpc region of the jets was found to be likely
oriented close to the plane of the sky. Analysis of the gradual
steepening of the spectral index along the jets and lobes provided 
an estimate of the radiative lifetime of $\le$ 160 Myr,
indicating that the source is likely very old.

The {\em XMM-Newton} observation of AWM\,4 revealed a very
relaxed system with no evidence of cooling in the cluster core and 
no X-ray cavities (O'Sullivan et al. 2005; Giacintucci et al. 2008;
Gastaldello et al. 2008). A deep $\sim$75 ksec {\em Chandra}
observation was recently presented by O'Sullivan et  al. (2010a, 
2010b). A small cool core (or galactic corona), with only $\sim$2 kpc
radius, was discovered at the group center, coincident with the radio
core. The cool core appears able to fuel the AGN for long
periods, possibly explaining the long outburst timescale of the radio 
source (O'Sullivan et al. 2010a). With the exception of a small X-ray 
surface brightness decrement at the center of the east lobe, no clear 
large X-ray cavities were detected at the location of the
lobes, suggesting that these are only partially filled by relativistic
plasma (O'Sullivan et al. 2010a). Analysis of the abundance
distribution in the cluster core revealed evidence of high-metallicity
material correlated with the radio jets (O'Sullivan et al. 2010b). The
jets may have entrained the gas enriched in the inner regions of the 
central galaxy and transported it out along their direction of propagation.

\subsection{NGC\,6269}\label{sec:ngc6269}

The cD galaxy NGC\,6269, at the center of the poor cluster of galaxies 
AWM\,5, hosts a low-power (Table 1), small radio source 
(Burns,  White \& Hough 1981; Giacintucci et al. 2007). 

\begin{figure*}
\centering
\epsscale{1}
\plotone{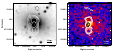}
\caption{NGC\,6269. {\em Left} -- {\em GMRT} 235 MHz contours
(FWHM=$14.1^{\prime\prime} \times 12.0^{\prime\prime}$, p.a. 
$80^{\circ}$; 1$\sigma$=600 $\mu$Jy beam$^{-1}$), overlaid on 
the POSS--2 red optical image. Contours are spaced by 
a factor 2 from $+3.0$ mJy beam$^{-1}$. Dashed contours correspond to
the $-3.0$ mJy beam$^{-1}$ level. 
{\em Right:} {\em GMRT} 610 MHz contours (FWHM=$5.3^{\prime\prime} 
\times 4.1^{\prime\prime}$, p.a. $69^{\circ}$; 1$\sigma$=70$\mu$Jy 
beam$^{-1}$), overlaid on the smoothed 0.3--2.0 keV {\em Chandra} 
image. Contours are spaced by a factor 2 starting from 0.28 mJy 
beam$^{-1}$. For this source the scale is 0.665 kpc/$^{\prime\prime}$.}
\label{fig:ngc6269}
\end{figure*}

Our {\em GMRT} images are shown in Fig.~\ref{fig:ngc6269}. On the left, the 235 MHz 
contours are overlaid on the optical image, while the right panel shows the 
X-ray emission of the cluster with the 610 MHz contours overlaid. The radio 
emission is totally contained within the optical envelope of the host galaxy.
In the $\sim$5$^{\prime \prime}$ resolution image at 610 MHz, the source is 
resolved in three components. Similar to the {\em GMRT} image at 1.28
GHz in Giacintucci et al. (2007), 
the axis of the central component is misaligned with respect to the
north-south axis 
connecting the outer lobes. 

The right panel of Fig.~\ref{fig:ngc6269} shows the smoothed {\em Chandra} image
of AWM\,5. The recent {\em Chandra} analysis by Baldi et al. (2009b) shows 
the system to possess a small ($\sim$ 8 kpc), dense cool core and two possible 
X-ray cavities at the locations of the radio lobes. However, these cavities are
difficult to study in detail because they are rather small and have a low
contrast with the surrounding emission.

\subsection{NGC\,7626}\label{sec:ngc7626}

NGC\,7626 is one of the two brightest ellipticals in the Pegasus galaxy group.
The dominant galaxy, NGC\,7619, lies $\sim100$ kpc from NGC\,7626
in the plane of the sky. A radio point-source is associated with NGC\,7619, while NGC\,7626 hosts a very 
extended radio source with symmetric jets (Jenkins 1982; Birkinshaw \& Davis 1985). 

\begin{figure*}
\centering
\epsscale{1}
\plotone{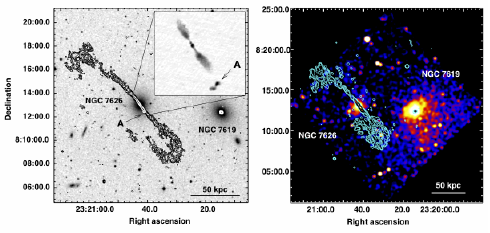}
\caption{NGC\,7626/NGC\,7619. {\em Left:} {\em GMRT} 610 full resolution contours (FWHM=
$6.2^{\prime\prime} \times 4.9^{\prime\prime}$, p.a. $31^{\circ}$; 
1$\sigma$=50 $\mu$Jy beam$^{-1}$), overlaid on the optical POSS--2 image. 
Contours are spaced by a factor 2 starting from 0.15 mJy beam$^{-1}$.
The inset shows the {\em VLA} 1.4 GHz image of the central region of NGC\,7626, 
with the resolution of 1.2$^{\prime \prime}$. {\em Right:} {\em GMRT} 235 MHz 
contours (FWHM=$14.2^{\prime\prime} \times 12.0^{\prime\prime}$, p.a. 
$57^{\circ}$; 1$\sigma$=800 $\mu$Jy beam$^{-1}$), overlaid on the 
smoothed 0.3--2.0 keV {\em Chandra} image. 
Contours are spaced by a factor 2 starting from 3 mJy beam$^{-1}$.
For NGC\,7626 the scale is 0.233 kpc/$^{\prime\prime}$.}
\label{fig:ngc7626}
\end{figure*}

Our {\em GMRT} images of the NGC\,7619/NGC\,7626 system are shown in Fig.~\ref{fig:ngc7626}
as contours on the POSS-2 and {\em Chandra} images. The high sensitivity of our 
observations allows us to image the WAT radio galaxy NGC\,7626 in its whole 
extent, including the low-surface brightness tails at the end of the jets. The total size 
of the source is $\sim$13$^{\prime}$, corresponding to approximately 180 kpc. 
A background radio source (A in the left panel) is projected onto the southern 
jet. The inset shows the 1.4 GHz image at a resolution of $1.2^{\prime \prime}$, 
obtained from the {\em VLA} archive (project AF188), which resolves the source into a small double. The
image also shows the bright core and inner jets of NGC\,7626. 
An unresolved radio source is detected at the position of NGC\,7619.
The flux densities of NGC\,7626 in Table 4 do not include the
contribution of the background radio
galaxy A, having a flux of 150$\pm$12 mJy and  62$\pm$3 mJy at 235 and
610 MHz, respectively.

The {\em Chandra} image in Fig.~\ref{fig:ngc7626} (right) shows the X-ray halo
of NGC\,7626 and the much larger and asymmetric X-ray emission associated
with NGC\,7619. This asymmetry might arise from gas stripped during the infall 
of NGC\,7619 into the group (Trinchieri, Fabbiano \& Kim 1997). The recent 
{\em Chandra} study by Randall et al. (2009) showed the presence of 
sharp X-ray brightness edges in each galaxies, which were interpreted as 
cold fronts generated by a major merger between the NGC\,7626 and NGC\,7619
subgroups. No substructures in the X-ray image appear to be associated 
with the extended radio emission from NGC\,7626, but there is an apparent
correlation between the southern lobe and the temperature map (Randall et 
al. 2009).

\section{Discussion}\label{sec:disc}

In this paper, we presented new low-frequency {\em GMRT} images at 235 MHz and 
610 MHz of 18 X-ray bright, nearby groups of galaxies.  Furthermore,
we carried out a qualitative comparison between the {\em GMRT} radio images and simple X-ray intensity
images from {\em Chandra} and {\em XMM-Newton} data to investigate the
presence of correlations between the radio structures and substructures in the
IGM. 

Table 5 summarizes the main radio and X-ray properties of the
groups. The number and type of X-ray cavities are from the literature, 
while the source class is defined in Sect.~6.4. The table shows that
our sample includes a large multiplicity of
situations both in the radio and X-ray bands. The radio sources are
extremely diverse in radio power, physical scale and morphology.
Powers range from very low-power FR I sources ($P_{\rm 235 \, MHz} \sim 5
\times 10^{21}$ W Hz$^{-1}$) to FR I/FR II transition
radio galaxies ($P_{\rm 235 \, MHz} \sim 10^{25}$ W Hz$^{-1}$),
and sizes range from galactic to group/cluster scales.  
The observed radio structures include classic, double-lobed radio galaxies,
both giant and extended on much smaller scale, tailed sources, 
morphologies similar to the core-halo sources found in cool core
clusters (e.g., Mazzotta \& Giacintucci 2008 and references therein) and more 
complex and irregular, extended structures.

In the X-ray band, the systems also appear very inhomogeneous. The X-ray
luminosities cover three orders of magnitude, i.e., from $L_X \sim 8
\times 10^{40}$ erg s$^{-1}$ (NGC\,1587) to $L_X \sim 4 \times
10^{43}$ erg s$^{-1}$ for the poor cluster AWM\,4. In half of
groups, the X-ray images show evidence of AGN-driven disturbances, the most spectacular examples
being cavities detected in the X-ray surface
brigthness (e.g., HCG\,62, NGC\,4636, NGC\,5813). Distorsions in the
group X-ray emission, edges and fronts are also observed in a number
of system; in some objects, these are
probably caused by the central radio source (e.g., NGC\,507), while in other sources
interactions between the group and a nearby, larger system might play a role
(e.g., NGC\,1407, NGC\,7626). The sample also includes examples of apparently 
relaxed groups, with no clear substructure in the X-ray surface
brightness, but with a significant level of disturbance in the temperature and 
abundance distribution (e.g, NGC\,3411).

The combined variety of radio and X-ray structures makes any classification 
of the systems in our sample difficult. The central radio
galaxy clearly perturbs the surrounding intragroup medium in some
groups, such as cavity systems. In other groups, the AGN/IGM interaction
does not strongly manifest itself in the X-rays, but rather 
through the effects of the high-density thermal gas on the morphology and
evolution of the radio-emitting plasma; the IGM can strongly distort
the radio structures or confine the radio plasma and prevent its
dissipation through adiabiatic expansion.

While a  detailed study of the radio/X-ray correlations in our group
sample will be presented in future papers (O'Sullivan et al. 2011, 
O'Sullivan et al. in preparation, Giacintucci et al. in
preparation), in the following sections we will provide a 
qualitative discussion of the effects of the IGM on the radio sources and
viceversa and derive some general conclusions on 
the interplay between the AGN and IGM for the groups in our sample.

\subsection{Radio power and linear sizes}
\label{sec:analysis}

The linear size of FR I radio galaxies is known to increase
with increasing radio power (Ledlow, Owen \& Eilek 2002, 
hereinafter referred as L02). No significant differences are found
between the size distribution, as a function of the power, for sources 
inside and outside the cores of rich clusters (L02).

\begin{figure}
\centerline{\includegraphics[width=8cm]{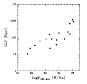}}
\caption{Radio power-radio size diagram for the group sample (NGC\,5846 was not observed at 235 MHz and therefore is not included
 in the plot). The empty circle and empty square in the plot on the left
mark the location of the complex radio sources in NGC\,5044
  and HCG\,15, respectively.}
\label{fig:plls}
\end{figure}

Using the 235 MHz powers and linear sizes in Table 5, we can plot the
radio largest linear size against $\log\, P_{\rm 235\ MHz}$ for our sources
(note that NGC\,5846 was not observed at 235 MHz and, therefore, is not included in the
analysis). The plot is shown in Fig.~\ref{fig:plls}. The empty circle
and empty square mark the location of the complex radio sources 
associated with NGC\,5044 (Fig.~\ref{fig:ngc5044}) and HCG\,15
(Fig.~\ref{fig:hcg15}), respectively. Our sources cover a wide range of spatial scale,
from LLS $\sim$10~kpc to over a Mpc (the largest extent being
in NGC\,315). Within this interval of
sizes, the radio power spans four orders of magnitude, from $P_{235 \, MHz}
\sim 5 \times 10^{21}$ W Hz$^{-1}$ (NGC\,4636) to $\sim 10^{25}$ 
W Hz$^{-1}$ (NGC\,383).

The size distribution of our groups seems in agreement with the
FR I power/size relation in L02, even though an accurate comparison 
is not possible (different frequencies and unspecified cosmology
used by L02). Although our sample is not a statistical sample, 
our result indicates that the
correlation between power and size is independent
of the environment of the galaxy,  as was 
suggested by the results of L02 for radio sources inside and outside 
clusters and by Giacintucci et al. (2007) for poor and rich
clusters. Our sources are located in similar environments, i.e., of optically
identified galaxy groups, yet we find among them a variety of sizes
and powers similar to that found for larger samples
of FR I sources which include rich cluster and field radio galaxies. 

\subsection{Radio morphology}\label{sec:morf}

The radio emission in our groups is characterized by a large variety
of structures (Table 5). In particular, the following morphologies 
were found:

\begin{itemize}

\item[1)]  three very extended ($> 500$ kpc) double sources, i.e., the well-known giant
sources B2\,0055+30 and 3C\,31 in NGC\,315 and NGC\,383, respectively, and the 
lobe-dominated double at the centre of MKW\,2. The 235 MHz observation of 3C\,31
allowed us to trace its emission on a total angular scale of $\sim$43$^{\prime}$ 
($\sim$ 900 kpc). This is the largest size reported so far for this
source. All three giant radio galaxies show disturbed and bent tails/lobes;

\item[2)] six double radio sources, with sizes between $\sim$6 kpc
  (NGC\,1407 at 610 MHz) and $\sim$200 kpc  (NGC\,7626). With the
  exception of UGC\,408, which shows two bright and straight 
jets surrounded by a diffuse cocoon, and NGC\,6269, a relatively symmetrical double with a bright 
central component, all the other doubles exhibit disturbed morphologies. These manifest themselves 
in the form of twisted and warped jets (NGC\,4636), strongly distorted and asymmetric 
lobes (NGC\,507), and bent lobes or tails (HCG\,62, NGC\,7626, NGC\,1407).
The small ($\sim 6$ kpc) asymmetric double, detected at 610 MHz
in NGC\,1407, is totally enclosed in a much larger ($\sim$80 kpc) emission
with a diffuse appearance that is detected only at 235
MHz. Such a peculiar radio structure might arise from restarted
radio activity (the small double) co-existing with the fading emission
from a former outburst (the diffuse component);

\item[3)] a $\sim$160 kpc wide-angle-tail in the poor cluster
  AWM\,4. The bent morphology of the source is likely caused by
the combined effects of ram pressure, caused by the motion southward of the host galaxy, 
and buoyancy forces (Giacintucci et al. 2008);

\item[4)] a bright nuclear source with a tiny ($\sim$3 kpc) extension
  at 610 MHz in NGC\,5846;

\item[5)] amorphous emission around the nuclear component in the
  groups NGC\,1587 and NGC\,3411, similar to the core-halo sources in 
cool-core clusters (e.g., Mazzotta \&  Giacintucci 2008 and
references therein), although on a much smaller scale;

\item[6)] very complex emission in the galaxy system NGC\,741/NGC\,742, including a
nuclear source in each galaxy,  a bridge connecting these components, 
a spectacular twisted $\sim$150 kpc long tail, and a region of diffuse emission to the east of 
NGC\,742. Such a peculiar morphology is likely caused by the superposition of two interacting 
radio galaxies. A possible interpretation is that the tail originates from NGC\,742, while 
the eastern diffuse structure is emission from NGC\,741, which is bent by the galaxy interaction;

\item[7)] diffuse, low-surface brightness emission with no centrally-peaked component and 
 extending for $\sim$150 kpc between the dominant galaxies of
 HCG\,15. Two of them, HCG\,15d (A) and HCG\,15a (B), host a point source;

\item[8)] a nuclear source and extended structure in NGC\,5044. The extended emission
has the form of a faint lobe at 610 MHz, which is not detected at 235
MHz.  At 235 MHz, the radio image reveals instead a distorted jet, misaligned with respect to the axis of the 610 MHz lobe,
and a detached region of low-surface brightness emission at $\sim$30
kpc from the nuclear component. Superposition of several radio structures
from multiple radio outbursts occurring at different times might
explain the morphological complexity of this system;

\item[9)] a possible double-double radio source $\sim$20 kpc size
  hosted by the giant elliptical NGC\,5813.

\end{itemize}

As for FR I radio galaxies in the dense
central regions of galaxy clusters (e.g., Owen, White \& Burns 1992; Owen, White \& Ge 1993; Giacintucci
et al. 2007 and references therein), distorted and asymmetric radio
morphologies are common for group-central radio galaxies. The
majority of our group sources show disturbances at some level.
The surrounding, dense thermal gas likely plays the major role in shaping
and deforming the extended components of radio sources, i.e., jets,
lobes and tails, both in groups and clusters.

\subsection{Radio spectral index and restarted activity}\label{sec:alpha}

The environment in which radio galaxies reside may affect not only the 
radio morphology, but also the spectral properties of the radio
emission. Radio galaxies in dense environments, such as cluster cores, 
tend to have steep radio spectra compared to objects in the field 
(e.g., Slee et al. 1983; Slee \& Reynolds 1984; Roland et al. 1985; Slee et
al. 2001; 	Bornancini et al. 2010). Cluster core radio sources have typical
spectral indices $\alpha>1$ up to extreme values of 
$\alpha \gtsim$ 2. Such steep radio spectra may be caused by the 
confinement of the relativistic plasma by the high pressure and dense external 
medium; the intergalactic medium prevents the radio source from
expanding and the evolution of the radio plasma is controlled
only by synchrotron and inverse Compton energy losses, 
resulting in a steepening of the radio spectrum.
Furthermore, the confinement of the relativistic plasma in dying
 radio sources -- objects where the nuclear activity has stopped and
the radio lobes are fading -- could sustain 
the radio emission longer than expected and prevent its fading
through adiabiatic expansion (e.g., Murgia et al. 2011 and references therein).

The spectral indices of 15 radio sources in our sample, measured
between 235 MHz and 610 MHz, are listed in Table 4. Only 6 of 15 
sources have $\alpha < 1$. All the remaining sources indeed have 
steep spectra, with $\alpha$ between $\sim 1.0$ and $\sim 1.4$ for 
the largest majority (6 of 9), and ultra-steep values for NGC\,5044
and NGC\,1407. An upper limit of
$\alpha < 1.9$ is listed for NGC\,315, as the flux density measured on 
the {\em GMRT} image at 610 MHz ($S_{610 \, MHz}=2.5$ Jy) is underestimated (see
Sect.~\ref{sec:ngc315}). Using the flux at 609 MHz measured by Mack et
al. (1998) on the {\em WSRT} image ($S_{609 \, MHz}=5.3$ Jy), the
spectral index of NGC\,315 becomes $\alpha=1.11$.

The result that 9/15 of the sources have a steep radio spectrum 
suggests that, as in clusters, group radio galaxies tend to have
steep radio spectra compared to objects residing in less dense
environments.

The ultra-steep spectra of NGC\,5044 and NGC\,1407 deserve 
further comment. Here, the steepness of the spectral index is determined by
the extreme discrepancy in structure and size between the radio
emission detected at the two frequencies, with the 235 MHz emission 
covering a much larger area  (one order of magnitude more extended in
the case of NGC\,1407) than the structure visible at 610 MHz. This
suggests the co-existence of at least two distinct radio outbursts in
both systems. The presence of old, steep-spectrum emission, generated
by former activity predominanting at 235 MHz, reflects the steep
value of $\alpha$ found for these two objects.

Similarly to NGC\,5044 and NGC\,1407,  there is evidence for 
multiple outbursts in HCG\,62 and NGC\,5813, and
possibly in NGC\,741 and NGC\,4636. The combination of the X-ray
properties (e.g., multiple cavities or multiple pairs of cavities)
and radio properties (e.g., steep-spectrum emission detected
only at 235 MHz or more extended emission at this frequency than 
at 610 MHz) suggests that restarted activity has occurred also in these groups.

Finally, NGC 507 represents a special case, as it has been argued that 
the source is a dying radio galaxy whose curved radio spectrum is 
dominated by the fading lobes (Murgia et al. 2011). 
The high pressure of the external
medium could have reduced or even
stopped the expansion of the lobes and the resulting confinement 
of the radio plasma might prevent the quick dissipation of the lobes 
through adiabiatic expansion.

NGC\,507 and the systems with restarted activity are all cases
where the IGM has strongly influenced the evolution of the
relativistic plasma. For instance, as in NGC\,507, the old, extended 
radio emission detected at 235 MHz in NGC\,5044 and NGC\,1407,  
with no counterpart at 610 MHz, will be visible at the lower frequency
because of the radio plasma confinement by
the dense X-ray emitting gas.

\subsection{Radio/X-ray comparison}

The targets of our {\em GMRT} project were chosen to show structures in
either the X-ray or radio which may indicate interactions between the
central radio galaxy and the intra-group medium. In half of the cases, we
did find` evidence for such interactions in the form of clear
associations between the radio and X-ray structures (e.g, radio-filled
X-ray cavities), but, in the remaining cases, the expected radio/X-ray
correlations appear less obvious or are not clearly detected.  In general,
the groups which were selected to possess significant substructure in the
X-ray band (mostly cavity systems) and with little radio emission at
frequencies $\gtsim$ 1 GHz, show radio structures in our deep,
low-frequency radio observations, which seem to correlate with the IGM
substructures (NGC\,5044, NGC\,5813, HCG\,62, NGC\,4636).  This may
imply that the X-ray emission preserves a record of inflated cavities for
a longer time than the high radio frequency emission, but not beyond the
time when the radio emission fades at lower frequencies.

Although a detailed classification of our systems is not
straightforward, we can group the sources into three broad classes 
(also reported in Table 5), which may help to elucidate the nature of 
the interaction occurring between the AGN and the IGM in these groups.

\paragraph{{\normalsize Class 1: Small radio sources with jet/lobe structures confined to
the group core.}}

In these groups we observe the strongest evidence of
AGN/IGM interactions. The most striking examples are the X-ray
cavity systems, which make up 10/18 of our groups (Table 5).  In 3
systems, cavities have a bi-polar distribution (HCG\,62, NGC\,5813,
NGC\,6269). In particular, HCG\,62 and NGC\,6269 possess a set of
cavities, symmetric with respect to the group center, both of which coincide with
the lobes of the central radio galaxy. NGC\,5813 represents an exceptional
case of multiple bi-polar cavities, with two pairs (inner and middle) of
radio-filled cavities and a possible third set of ghost cavities, all
nicely aligned along the same axis.  In two systems (NGC\,5044 and
NGC\,4636), cavities appear almost isotropically spread around the centre at
roughly the same radius, although they must have been produced at
different times.  The fact that some of these cavities are filled by radio-emitting
plasma, while others are radio-quiet even at low frequencies, might
reflect the different ages of the cavities. Group weather, due to
previous AGN outbursts or motion of the galaxy with respect to the 
center of the group potential, may be responsible for the isotropic
distribution of the cavities (David et al. 2009, 2011).

In the radio band, all cavity systems, with the exception of the ghost
cavity in NGC\,741 and the small cavity in the east lobe of AWM\,4, 
host a relatively small, low- or mid-power double radio source (Table
5). While this could be biased by the difficulty in
detecting X-ray cavities at larger radii due to the lower surface brightness, it may
also suggest that cavities preferentially occur when we have a less
powerful source expanding into the relatively dense environment of
a group core.

\paragraph{{\normalsize Class 2: Large radio sources with jet/lobe structures extending
beyond the group core.}} 

This category includes the groups with the powerful,
extended radio sources NGC\,315, NGC\,383, MKW\,2, and NGC\,7626, 
whose radio lobes/tails escape from the group core and reach a large
distance (between $\sim$90 and $\sim$600 kpc)
from the center. No clear cavities are detected at the lobe/tail location in
the current X-ray images of these systems and in some cases the radio
source extends outside the X-ray field of view. However, we cannot exclude the
possibility that cavities are present, but are undetected due to the low
X-ray surface brightness in such a peripheral region of the group X-ray
halo. Cavagnolo et al. (2010) categorize some of these systems as
"poorly confined" and suggest that their large sizes indicate faster
moving jets which have entrained only small masses of gas from the group core.

\paragraph{{\normalsize Class 3: Amorphous/diffuse radio sources.}} 

The groups NGC\,1587 and NGC\,3411 are characterised by 
nuclear radio emission from the central galaxy surrounded by an amorphous
component, rather than distinct jets or
lobes. Their sizes are rather small ($\sim$20 kpc for NGC\,1587 and 80
kpc for NGC\,3411) and they have
low to moderate radio powers (Table 5). These groups do not
show X-ray cavities and the lack of radio jets and lobes seems to
naturally
explain their absence. 

Groups may also host diffuse radio emission that is not clearly associated with individual group
galaxies, but rather with the intra-group medium. This is the case of
HCG\,15, a system that does not appear to be completely relaxed,
and contains several bright ellipticals rather than a
single dominant galaxy.  While such radio emission might be the remnant
of an old, detached radio lobe subject to the intra-group gas motion, it may
also be emission arising from a broad shock front generated by the passage
of a galaxy through the core of the group, as suggested by the comparison
with Stephan's Quintet (O'Sullivan et al. 2009).
\\
\\
We note that some systems can be placed in more than one of the
classes described above.  For instance AWM\,4, NGC\, 7626 and
NGC\,741 are objects between classes 1 and 2, having extended radio jets
or tails (LLS$\sim 200$ kpc) and some indication of cavities. Similarly,
NGC\,1407 can be classified as class 1 or 3 depending on the radio
frequency: the group hosts a small double at 610 MHz (class 1), but, at
the same time, the detection of diffuse emission at 235 MHz places the source
in class 3. Overlaps of this sort are inevitable in any
classification scheme, but we consider them to be informative rather
than an indication that the scheme is over-simple.

The temperature structure of the groups will be addressed in a
future paper, but in broad terms we find that systems in classes 1 and 2
typically have cool cores, while some of those in class 3 do not, at least
at the spatial resolutions achievable with the available X--ray data. This
is in agreement with the finding of Sun (2009) that group and cluster
dominant galaxies with strong radio AGN are located in a cool core.  The
physical scale of the cool cores varies between groups, with some of our
systems having small-scale cool cores or galactic coronae. We note that
these occur in both classes 1 and 2, e.g., NGC\,6269 (Baldi et al. 2009)
and AWM\,4 (O'Sullivan et al. 2010a).

\subsection{Gas entropy excess in galaxy groups}

Poor clusters and groups show a significant excess of gas entropy with
respect to the expectations from gravitational interaction of gas
with dark matter (Finoguenov et al. 2002, Ponman, Sanderson \& Finoguenov 2003, and
references therein). This is interpreted as evidence for
non-gravitational processes affecting the properties of the IGM.
The entropy enhancement could occur at high redshift, by heating 
the gas at or before the time of group formation
(pre-heating). Alternatively, the entropy might be raised at low redshift by
heating the gas in the core or by removal of the coolest gas.

AGN feedback is now widely regarded as the most likely 
mechanism for increasing entropy at low redshift. Numerous studies 
have shown that the energy stored in radio lobes and X-ray cavities is
sufficient to balance radiative cooling across a wide range of mass 
scales (see McNamara \& Nulsen 2007, and references therein).
At group scales, the idea seems to find support in the tentative 
evidence that the IGM in radio-loud groups has on average a higher 
temperature than radio-quiet groups of the same luminosity 
(Croston, Hardcastle \& Birkinshaw 2005). Studies of the
two-dimensional entropy distribution of galaxy groups also
suggest that episodic heating, rather than passive evolution after an
early preheating phase, is required to reproduce observed IGM properties
(Johnson, Ponman and Finoguenov 2009). On the other hand,  
Jetha et al. (2007) find that the entropy profiles of groups with a 
central bright radio source 
are not significantly different from those of radio-quiet groups. 
The Jetha et al. (2007) analysis focusses on the central regions of
groups, whereas Croston et al. (2005) use temperatures derived 
from a larger fraction of the group halo, which may in part explain 
the difference. Furthermore, Croston et al. (2005) use radio data at 1.4 GHz, which 
are mostly sensitive to the current phase of activity of the AGN, but
in fact the increased entropy in some groups may be driven by past AGN activity.
In some AGN heating models, the entropy excess can indeed last for a 
considerable long time after the AGN activity has ceased
(Roychowdhury, Ruszkowski \& Nath 2005). Dwarakanath \& Nath 
(2006) approached this issue by selecting from the sample in Croston 
et al. (2005) radio-quiet groups that most deviate from the
extrapolation of the $L_X-T$
relation from rich clusters and that do not show signs of current AGN
activity at 1.4 GHz that could explain such a deviation. In search 
of radio emission produced by possible past AGN activity, they
observed these systems with the {\em GMRT}  at lower frequency, 
but found no evidence of such radio emission.

Numerical simulations including AGN feedback find that strong 
feedback can raise the entropy of the IGM to a level comparable to
what is observed (McCarthy et al. 2010; Muanwong et al. 2002). In these
AGN heating models, most of the work of AGN feedback occurs
at fairly high redshift, when the energy input from the AGN is sufficient to
remove gas from low mass groups (less than a few times 10$^{14}$
$M_{\odot}$) or, alternatively, to increase the entropy of the gas so
much that this material will never be accreted by the system. 

This raises the interesting question whether the examples of possible 
AGN feedback seen in our sample of local groups are in conflict with these
models. The nature of our sample, which by selection includes gas
rich systems where the signs of the AGN/IGM interaction can be seen 
in the X-ray images, makes it difficult to answer this question. If
high-redshift heating has removed a considerable
amount of gas from groups, these systems would be too faint for detailed observations in the
X-ray band and, thus would not fall within our selection.
The outbursts in most of our systems are not powerful enough to
eject gas from the group. Two of the most powerful radio sources in
the sample, NGC\,315 and NGC\,383, reside in groups which are gas poor. 
However, the mismatch 
between source size and area covered by the X-ray image prevents a
clear understanding of how (and if) the radio source is affecting the
intra-group medium in these two systems.  In half of the sample we observe
X-ray substructures likely induced by AGN activity, suggesting that
AGN outbursts may be affecting the gas entropy right now. In
particular, cavities buoyantly rising from the group center should
temporarily alter the entropy profile of the gas by uplifting gas behind
them, and shocks (as, for instance, in NGC\,5813, Randall et al. 2011) 
will increase the entropy of the gas. The systems in the sample 
with evidence of AGN/IGM interaction are mostly gas-rich groups 
with low- and intermediate-power central 
radio galaxies. This suggests that the current AGN
activity in these systems is weak compared to that proposed as the cause of
gas expulsion at high redshifts, and perhaps that our sample is biased toward
systems which had less AGN heating at high redshift and so retained most of their gas.

A more detailed analysis of the gas properties of our sample is
deferred to a future paper and 
may provide valuable information on the way gas entropy in groups
is altered by AGN activity in the local Universe. Because of the bias
toward gas rich groups, our sample is probably not suitable to test the
proposed AGN heating models at high redshift.

\section{Summary}\label{sec:summary}

In this paper, we presented low-frequency {\em GMRT} observations of 
18 nearby ($z < 0.05$), elliptical-dominated groups of
galaxies. The targets were chosen to possess structures, either in the
X-ray surface brightness and temperature maps or radio morphology
from the literature and radio archives, which may indicate interaction
between the radio source and the surrounding intra-group medium. 

The {\em GMRT} observations were carried out at 235 MHz and 610 MHz, 
with a typical observing time of $\sim$2-3 hours on source and a
sensitivity in the final images in the range 35-100 $\mu$Jy at 610 MHz 
and 0.2-1 mJy at 235 MHz. Our main results are summarized below.

\begin{itemize}

\item[1.] The {\em GMRT} images show a very diverse population of
radio sources associated with the central elliptical galaxy in the
groups. Similarly to FR I radio sources in galaxy clusters, our sample
includes classic double-lobed sources, both giant and extended on the
galactic scale, wide-angle-tail and head-tail
sources, amorphous sources with a nuclear component resembling the
core-halos in cool-core clusters (although on a much smaller scale), 
and more complex and irregular morphologies. In the unrelaxed group
HCG\,15, we detected low-brightness, diffuse emission extended 
on the group scale and not clearly associated with any of the group
galaxies.

\item[2.]  The radio powers span four orders of magnitude, from $10^{21}$ W  
Hz$^{-1}$ (i.e., very low-power FR I radio galaxies) up to 10$^{25}$ W Hz$^{-1}$
(i.e., FR I/FR II transition radio galaxies). The linear sizes cover a
wide range of spatial scales, from LLS$\sim$10 kpc to $\gtsim$1 Mpc.
The distribution of the linear size as a function of the radio power
is in general agreement with the power/size correlation 
of FR I radio galaxies inside and outside the cores of rich clusters
(L02). This may suggest that the correlation between power and size
is not dependent on the environment in which the radio galaxies reside.

\item[3.] The comparison of the {\em GMRT} images with the {\em Chandra} and {\em
XMM-Newton} images of the group X-ray emission reveals associated radio/X-ray
emission in approximately half of the groups, the most evident
example being the radio-filled X-ray cavities. The remaining systems
show little or no clear correlation in the radio and X-ray images.
In general, those groups with strong substructure in the X-ray images
and with weak radio emission at $\gtsim$ 1 GHz, show radio structures
in our low-frequency {\em GMRT} images, which correlate with the IGM 
substructures (NGC\,5044,  NGC\,5813, HCG\,62, NGC\,4636).  
This suggests that the X-ray emission preserves a record of inflated cavities for a
longer time than the high radio frequency emission, but not beyond the
time when the radio emission fades at lower frequencies.

Although possibly biased by the difficulty in detecting X-ray cavities
in large radio sources with jets/lobes extending beyond the group core
(NGC\,315, NGC\,383, MKW\,2, NGC\,7626), cavity systems are typically
associated with small double radio
sources with a low or medium radio power 
(e.g., HCG\,62, NGC\,6269). Multiple cavity systems, with both
radio-filled and ghost cavities, are observed in NGC\,5813, NGC\,4636 
and NGC\,5044. Here, repetitive radio outbursts could have inflated
sets of cavities at different epochs. All multiple pairs of cavities have a bipolar
distribution and are aligned along the same axis in NGC\,5813, while 
{\em group weather} (David et al. 2011) could have spread the cavities isotropically around the 
group center in NGC\,4636 and NGC\,5044.

The lack of jets and lobes in amorphous radio sources such as
NGC\,1587, NGC\,3411 and HCG\,15, provides a natural
explanation for the absence of X-ray cavities in these type of systems.

\item[4.] The AGN/IGM interaction does not only manifest itself via 
evident radio/X-ray correlations, but also through the consequences of
the high pressure of the IGM on the radio plasma and its evolution. 
We found that, as for cluster FR I radio galaxies, both the radio structures and 
spectral properties of our group-central sources appear strongly affected by the
surrounding, dense thermal gas. The majority of the radio sources 
show morphological disturbances at some level, likely caused by the
interaction with the IGM, which has shaped and deformed the radio jets, 
lobes and tails. A steep radio spectrum ($\alpha > 1$) is observed in
$\sim 60 \%$ of our sources, suggesting that, as in clusters, radio
galaxies in the group cores tend to have steep radio spectra compared
to objects in less dense environment. Such steep spectra may be explained
in terms of the confinement of the relativistic plasma by the IGM,
which can reduce and even inhibit the expansion of 
of the radio source, leaving only synchrotron and inverse Compton
energy losses to control the evolution of radio plasma, resulting
in a steepening of the spectral index.

The radio and/or X-ray properties provide evidence of at least two
distinct radio outburts in NGC\,1407 and NGC\,5044 and
hint at recurrent radio activity in other four groups of the sample
(HCG\,62, NGC\,5813, NGC\,4636, NGC\,741). 
In these cases, the co-existence of the emission from fresh activity and from
former outbursts is (wholly or partially) responsible for the
steepness of the measured radio spectra.

The radio plasma confinement by the IGM can also sustain the radio
emission for long times and prevent the extended emission in 
old, dying radio galaxies (e.g., NGC\,507), and those with multiple
outbursts, from fading through adiabatic expansion. For instance, the confinement may 
allow the low-frequency detection of steep-spectrum radio emission 
associated with former activity in restarted radio galaxies such as 
NGC\,1407, NGC\,5044 and NGC\,5813.

\end{itemize}

In future papers (O'Sullivan et al. 2011, O'Sullivan et al. and
Giacintucci et al. in preparation), we will compare the radio and X-ray properties of
the groups in more detail, derive the spectral properties (integrated spectra and
spectral index images), physical parameters and ages of the radio
sources, and investigate their interaction with the group environment.

\acknowledgments 
We thank the anonymous referee for constructive comments that 
improved this work. We thank the staff of the {\em GMRT} for their
help during the observations.
{\em GMRT} is run by the National Centre for Radio Astrophysics of the Tata
Institute of Fundamental Research. This research has made use of the
NASA/IPAC Extragalactic Database (NED), which is operated by the Jet
Propulsion Laboratory, California Institute of Technology. SG
acknowledges the support of NASA through Einstein Postdoctoral
Fellowship PF0-110071 awarded by the {\em Chandra}
X-ray Center (CXC), which is operated by SAO. EOS acknowledges 
the support of the European Community under the Marie Curie Research Training Network.
JMV, LPD, CJ and WRF acknowledges the support of the Smithsonian 
Institution and CXC. This work was supported by grants
GO8-9127XR, AR0-11017X and GO0-11003X issued by the {\em Chandra} X-ray
Observatory Center, {\em XMM-Newton} grant NNX07AE95GR and
grant ASI-INAF I/088/06/0. Basic research in astronomy at the Naval 
Research Laboratory is supported by 6.1 base funding.
This research has made use of data obtained from the {\em Chandra} and
{\em XMM-Newton} Data Archives,  and software provided by CXC in 
the application packages CIAO, ChIPS, and Sherpa.

\clearpage



\end{document}